\documentclass[a4paper,fleqn,usenatbib]{mnras}

\usepackage{newtxtext,newtxmath}
\usepackage[T1]{fontenc}
\usepackage{ae,aecompl}
\usepackage{graphicx}	
\usepackage{amsmath}	
\usepackage{bm}
\usepackage[dvipsnames]{xcolor}
\usepackage[normalem]{ulem}

\newcommand{\nb}{n_{\rm b}}
\newcommand{\nc}{n_{\rm c}}
\newcommand{\ns}{n_{\rm s}}
\newcommand{\prob}{{\rm P}}
\newcommand{\normal}{{\rm{N}}}
\newcommand{\uniform}{{\rm U}}
\newcommand{\dirichlet}{{\rm D}}
\newcommand{\invwish}{{\rm W}^{-1}}
\newcommand{\alphas}{{\bm a}}
\newcommand{\specmean}{{\bm m}}
\newcommand{\speccov}{{\bm S}}
\newcommand{\classprob}{{p}}
\newcommand{\classprobs}{{\bm p}}
\newcommand{\objspec}{{\bm s}}
\newcommand{\objclass}{{\kappa}}
\newcommand{\objclasses}{{\bm \kappa}}
\newcommand{\objdata}{\hat{\bm d}}
\newcommand{\objnoise}{{\bm N}}
\newcommand{\identity}{{\bm I}}
\newcommand{\scalemat}{{\bm \Gamma}}
\newcommand{\wfmean}{{\bm w}}
\newcommand{\wfcov}{{\bm W}}
\newcommand{\condcov}{{\bm C}}

\title[Stellar Spectra as Sparse Non-Gaussian Processes]{SSSpaNG! Stellar Spectra as Sparse, data-driven, Non-Gaussian processes}
\author[S. M. Feeney et al.]{
Stephen M. Feeney,$^{1,2}$\thanks{E-mail: stephen.feeney@ucl.ac.uk}
Benjamin D. Wandelt,$^{1,3,4}$
and Melissa K. Ness$^{1,5}$ 
\\
$^{1}$Center for Computational Astrophysics, Flatiron Institute, 162 Fifth Avenue, New York, NY 10010, USA\\
$^{2}$Department of Physics and Astronomy, University College London, London WC1E 6BT, UK\\
$^{3}$Sorbonne Universit\'e, CNRS, UMR 7095,  Institut d'Astrophysique de Paris (IAP), 98 bis boulevard Arago, 75014 Paris, France\\
$^{4}$Sorbonne Universit\'e, Institut Lagrange de Paris (ILP), 98 bis boulevard Arago, 75014 Paris, France\\
$^{5}$Department of Astronomy, Columbia University, Pupin Physics Laboratories, New York, NY 10027, USA
}

\date{Accepted XXX. Received YYY; in original form ZZZ}

\pubyear{2019}

\begin{document}
\label{firstpage}
\pagerange{\pageref{firstpage}--\pageref{lastpage}}
\maketitle

\begin{abstract}
Upcoming million-star spectroscopic surveys have the potential to revolutionize our view of the formation and chemical evolution of the Milky Way. Realizing this potential requires automated approaches to optimize estimates of stellar properties, such as chemical element abundances, from the spectra. The sheer volume and quality of the observations strongly motivate that these approaches should be driven by the data. With this in mind, we introduce SSSpaNG: a data-driven non-Gaussian Process model of stellar spectra. We demonstrate the capabilities of SSSpaNG using a sample of APOGEE red clump stars, whose model parameters we infer using Gibbs sampling. By pooling information between stars to infer their covariance, we permit clear identification of the correlations between spectral pixels. Harnessing this correlation structure, we infer the true spectrum of each red clump star, inpainting missing regions and denoising by a factor of at least two for stars with signal-to-noise of $\sim$20. As we marginalize over the covariance matrix of the spectra, the effective prior on these true spectra is non-Gaussian and sparsifying, favouring typically small but occasionally large excursions from the mean. The high-fidelity inferred spectra produced with our approach will enable improved chemical elemental abundance estimates for individual stars. Our model also allows us to quantify the information gained by observing portions of a star's spectrum, and thereby define the most mutually informative spectral regions. Using 25 windows centred on elemental absorption lines, we demonstrate that the iron-peak and alpha-process elements are particularly mutually informative for these spectra, and that the majority of information about a target window is contained in the 10-or-so most informative windows. Such mutual information estimates have the potential to inform models of nucleosynthetic yields and the design of future observations. Our code is made publicly available at \href{https://github.com/sfeeney/ddspectra}{https://github.com/sfeeney/ddspectra}. 
\end{abstract}

\begin{keywords}
keyword1 -- keyword2 -- keyword3
\end{keywords}


\section{Introduction}
\label{sec:intro}

Surveys such as APOGEE \citep{Majewski2017}, GALAH \citep{deSilva2015}, Gaia-ESO \citep{Gilmore2012}, LAMOST \citep{Newberg2012}, SEGUE \citep{Yanny2009} and RAVE \citep{Steinmetz2006} have provided a vast dataset of spectroscopic observations that has revolutionized our view of the Milky Way, through corresponding velocity, stellar parameter, individual abundance and age measurements~\citep[e.g.][]{Nidever2014,Minchev2014a,Hayden2015,Kord2015,Ho2017,Frankel2018,Bovy2019,Ted2019a,BH2019}. In the coming years, large spectroscopic surveys such as Sloan V \citep{Kollmeier2017}, WEAVE \citep{Bonifacio2016}, 4MOST \citep{deJong2016}, PFS \citep{PFS2016}, Gaia RVS \citep{Gaia2016} and MOONS \citep{C2014} will begin observations, expanding the spectral data we have collected for our Galaxy by orders of magnitude. At present, the large ($>$ 10$^5$ star) medium-resolution surveys, such as APOGEE (R=22,500), rely on expensive observations, integrating to signal-to-noise ratios (SNRs) of up to 100 per pixel~\citep{Zasowski2013,Zasowski2017}. 

High-SNR spectra have been often regarded as necessary in the pursuit of precision abundances, required for  chemical differentiation across the Galaxy.  These abundances trace the detailed chemical evolution of the Milky Way, which is driven by an ensemble of stellar explosion and mass-loss activity. In the Galactic disk, where the majority of the stellar mass resides, abundances provide~\citep{Rix2013,BH2016} the record of its inside-out formation over time. The earliest epoch of the Galaxy's formation and its continued interaction with its environment, is documented in the chemical composition and characteristics of the stellar halo~\citep[e.g.][]{Keith2015,Payel2019, Helmi2018}.  Current data place strong constraints on the chemical evolution models designed to explain Galactic formation and evolution~\citep[e.g.][]{Minchev2013,Minchev2014b,Robyn2018, Weinberg2019, Clarke2019, Blancato2019}. Upcoming data offer the opportunity to refine these models considerably: for example, the disk is also believed to comprise  numerous individual birth sites where groups of stars were born. Any prospect of assigning stars to their birth sites via their unique chemical signatures~\citep[e.g.][]{BH2010} requires large stellar numbers and high precision abundance measurements~\citep{Mits2013, Ting2015, Hogg2016, Arm2018}. 

The large data volumes now in hand have led to the development of new approaches for deriving abundance measurements from spectral data, driven by the need for automatic, efficient means of extracting the full information content of the data. These include data-driven modeling approaches such as The Cannon \citep{Ness2015}, full spectral fitting using physical models as implemented in The Payne \citep{Ting2018} and  deep learning \citep{Leung2019}. These approaches improve the precision of abundance measurements significantly, permitting useful abundances to be estimated using 1/4 to 1/9th of the observing time compared to previous approaches. Specifically, abundance precisions on the order of 0.05-0.1 dex can be achieved at SNR $\approx$ 40 per pixel \citep{Ho2017b,Ness2018,Ting2018, Leung2019}. It has also been demonstrated that an ensemble of individual abundances can be derived at medium (R=11,000) and low (R=1,800) resolution by full spectral modeling \citep[e.g.][and Wheeler et al., in prep.]{Casey2016,Ting2017}. Physically, this is well-justified: abundances can be measured from their impact on the entire spectral range as legitimately as from individual elemental lines~\citep[e.g.][]{Ting2018}. This methodological advance in particular is relevant for the Gaia RVS data (R=11,000 spectra for 7 million objects) and, furthermore, the large ensemble of low-resolution data being observed in future surveys. The dramatic and rapid increase in available spectra and availability of increasingly powerful computational resources means we find ourselves in an era of tremendous opportunity for developing new avenues of stellar spectral modeling.

Central to the success of The Cannon and The (Data-Driven) Payne \citep{Xiang2019} is pooling: sharing information between members of a population to improve our knowledge of individual stars. In The Cannon, pooling is performed in a data-driven fashion by learning the relationship between stellar spectra and individual stellar abundances; in The Payne, (during the training step) by calibrating physical models of stellar spectra using labels derived therefrom. In this paper, we seek to generate a data-driven model of the stellar spectra themselves, as opposed to the abundance measurements, formalizing this concept of pooling within a Bayesian hierarchical model~\citep{Gelman_etal:2013}. By sharing information between stars, we will generate more precise representations of individual spectra, directly infer the correlation structure between spectral pixels and, in the process, gain understanding of the information content of the data. To date, there has been little work on the characterization and interpretation of the correlations between (and the dimensionality of) spectral data \citep[see however][]{Ting2012, PJ2019, M2014, Casey2019}. Our methodology will provide a direct measure of the information content of spectral regions and, correspondingly, elemental abundances.
 
We use stars observed by the APOGEE survey to build an extremely general and flexible empirical model of a large set of spectral data. Specifically, we implement a Gaussian Process~\citep{Rasmussen_Williams} mixture model representation of the APOGEE red clump stars. Unlike typical Gaussian Process analyses, we infer each element of our covariance matrices directly, without assuming a kernel function, and marginalize over the covariances when quoting our inferred true spectra. As a result, and contrary to analyses in which the covariance is fit once and fixed, the prior distributions of our true spectra are highly non-Gaussian, with a sparsifying prior whose negative logarithm is non-convex. Our method is a significant new technical advance in the modeling of stellar spectra and is distinct from, but builds upon, existing progress in data-driven spectral modeling in the regime of large data sets. We use no physical knowledge in constructing our model or selecting priors, and our inference is therefore entirely driven by common trends in the high-dimensional APOGEE data. In successfully pooling information about stars we achieve the following for the APOGEE spectra:

\begin{enumerate}
\item Prediction of masked (unmeasured or contaminated) regions of the spectra to enable, e.g., abundance estimates that would otherwise be impossible (see Sections~\ref{sec:validation} and~\ref{sec:inference}). This is particularly valuable in APOGEE for neutron-capture elements such as Nd and Ce, for which only a handful of weak features exist from which to estimate abundances. Some of these elemental features may fall near one of three chip gaps and therefore be absent in some (but, critically, not all) spectra due to stellar velocities. Our modeling of the data can predict these regions when they are absent. 
\item Denoising of all spectra, enabling higher precision inference at lower SNR (see Section~\ref{sec:inference}). The utility of this feature depends on the size of the effects we wish to discover. Our expectation is that this is particularly useful for weak features on the limits of detection, similar to previous demonstrations using generative modeling (e.g. The Cannon and The Payne).
\item Detailed examination of the empirical correlations in the spectra, quantitative measurements of these correlations and identification of which elemental absorption lines are positively and negatively correlated (see Section~\ref{sec:inference}).
\item Quantification of the information content of the data and determination of the most informative regions of spectra (see Section~\ref{sec:info}). This has consequences for both theory and experimental design. Along with the correlation structure we infer, the information content that we measure should place strong constraints on physical models of nucleosynthesis and chemical evolution. From an experimental design perspective, quantifying the informativeness of regions of spectra can drive the selection of wavelength ranges optimized for specific scientific purposes, answering questions such as whether we can retain sensitivity to abundances by observing a reduced spectral range, or conversely whether we gain significant information on a range of elemental features by observing a particular set of wavelengths.
\end{enumerate}

In the following, we describe the APOGEE data we use in Section~\ref{sec:data} and our model and its inference in Section~\ref{sec:methods}. We present our results in Section~\ref{sec:results} and discuss their consequences, current limitations and plans for their resolution in Section~\ref{sec:discussion}.


\section{Data}
\label{sec:data}

For our modeling we use the APOGEE red clump spectra from DR14 \citep{Majewski2017, Bovy2014}. These spectra comprise $29502$ stars with a mean SNR of 210 and range of SNR of 21-1775.  The contamination of red giant branch stars within this sample is on the order of 5-10 percent \citep{Bovy2014}. While our approach is applicable to any stellar population, we select a largely homogeneous population for this proof of concept, restricting to the narrow temperature and gravity range of the red clump stars. Doing so should reduce the number of components required for our Gaussian Process mixture model, simplifying its inference.

The data have been downloaded from the APOGEE database having already been shifted in radial velocity back to the rest frame and continuum-normalized \citep[see][]{Nidever2015}, with a slight SNR dependence on the continuum normalization that we discover with our Gaussian Process modeling. The spectra cover the range 15100.80-16999.81 \AA, and comprise $\nb = 8575$ spectral bins. Repeated inversion of the $\nb \times \nb$ covariance matrices required for inference would be prohibitively slow, and we thus restrict our analysis to a set of 25 spectral windows centred on lines confidently assigned to 25 different individual elements. These element windows have been chosen from the set of windows used to drive the APOGEE abundances in consultation with Jon Holtzman and Matthew Shetrone \citep{Holtzman2015, Shetrone2015}. Specifically, we process all spectral bins within $\pm 1.5$ \AA\ of the line centres specified in Table~\ref{tab:window_centres}, reducing the number of spectral bins to $\nb = 343$ and hence inversion time by a factor of $\sim15000$. The elements responsible for these absorption lines can be grouped into the following nucleosynthetic families: iron-peak, alpha-process, r-process, s-process, light and light with odd atomic number. We expect that common production mechanisms should correlate elemental abundances and hence these spectral windows. To examine correlations between and within the nucleosynthetic family members, we colour the elements by their families in relevant figures throughout the paper, setting out these colours in Table~\ref{tab:window_centres}.

In selecting the windows to examine we were confronted with a series of choices, each of which ultimately impacted the emphasis of our downstream analysis. Given the potential computational expense of this modeling approach, for our proof-of-concept analysis we adopted only a single line region for every element, but for as many elements as possible, thereby prioritizing breadth across elements in our correlation and inpainting investigations. The requirements for a line to be selected were that it be identified as one of the windows used by APOGEE in their processing pipeline ASPCAP \citep{GP2015}, as well as being both strong and, where possible, unblended. In some cases, multiple lines fitted these criteria for a single element.

\begin{table}
    \centering
    \caption{The list of the 25 elements that we select for our spectral modeling and their corresponding central wavelength (in a vacuum) corresponding to Figure 4.}
    \label{tab:window_centres}
    \begin{tabular}{lcl}
        \hline
        element & window centre / \AA & elemental family \\
        \hline
        Al & 16723.500 & \textcolor{OliveGreen}{light odd-Z (green)} \\
        C & 15582.101 & \textcolor{NavyBlue}{light (blue)} \\
        Ca & 16155.176 & \textcolor{red}{alpha (red)} \\
        Ce & 15789.063 & \textcolor{Sepia}{s-process (brown)} \\
        Co & 16158.700 & \textcolor{orange}{iron peak (orange)} \\
        Cr & 15684.264 & \textcolor{orange}{iron peak (orange)} \\
        Cu & 16010.023 & \textcolor{orange}{iron peak (orange)} \\
        Fe & 15495.100 & \textcolor{orange}{iron peak (orange)} \\
        Ge & 16764.238 & \textcolor{Sepia}{s-process (brown)} \\
        K & 15167.081 & \textcolor{OliveGreen}{light odd-Z (green)} \\
        Mg & 15745.000 & \textcolor{red}{alpha (red)} \\
        Mn & 15221.867 & \textcolor{orange}{iron peak (orange)} \\
        N & 15321.871 & \textcolor{NavyBlue}{light (blue)} \\
        Na & 16378.276 & \textcolor{OliveGreen}{light odd-Z (green)} \\
        Nd & 15372.342 & \textcolor{Plum}{r-process (purple)} \\
        Ni & 15559.517 & \textcolor{orange}{iron peak (orange)} \\
        O & 15760.300 & \textcolor{red}{alpha (red)} \\
        P & 15715.930 & \textcolor{OliveGreen}{light odd-Z (green)} \\
        Rb & 15293.534 & \textcolor{Sepia}{s-process (brown)} \\
        S & 15482.319 & \textcolor{red}{alpha (red)} \\
        Si & 15964.600 & \textcolor{red}{alpha (red)} \\
        Ti & 15339.241 & \textcolor{red}{alpha (red)} \\
        V & 15929.052 & \textcolor{orange}{iron peak (orange)} \\
        Y & 15624.142 & \textcolor{Sepia}{s-process (brown)} \\
        Yb & 16502.973 & \textcolor{Plum}{r-process (purple)} \\
        \hline
    \end{tabular}
\end{table}

\section{Methods}
\label{sec:methods}

Gaussian processes are a conceptually simple yet extremely powerful tool for regression and classification~\citep{Rasmussen_Williams}. Put briefly, a Gaussian process is a set of random variables whose joint distribution is multivariate normal, and is therefore fully specified by a mean function and covariance function. By their (Gaussian) nature, Gaussian processes permit simple, often analytically tractable, inference of their mean and covariance functions given (potentially noisy) observations, yielding flexible non-parametric fits to underlying trends in data and probabilistic predictions for new observations. As a result, Gaussian processes have found use throughout astronomy, from cosmology \citep{Bond_etal:1987} and cosmography \citep{Shafieloo_etal:2012} to models of instrumental systematics \citep{Gibson_etal:2012}, exoplanet populations \citep{DFM_etal:2014} and stellar spectra \citep{Czekala_etal:2017}.

In this work, we model the underlying ``true'' spectrum ($\objspec_i$), of the $i^{\rm th}$ APOGEE red clump star as a draw from a Gaussian process with a mean spectrum ($\specmean$) and covariance ($\speccov$) to be inferred from the data. In typical Gaussian Process models, the covariance function is taken to be one of a number of standard kernels \citep{Rasmussen_Williams}, chosen to reflect known or assumed properties of the observation and/or physical process (e.g., stationarity, isotropy, or periodicity). In the following, we do not assume an analytic form for our covariance function as is traditional in Gaussian Process models. Rather, we infer the correlations between the observed spectral bins, i.e., the individual elements of the covariance matrix. By doing so, we remove any potential for bias induced by a sub-optimal kernel choice incorrectly enforcing stationarity, a single correlation length, or a particular line shape, for example. As a result, we can not make predictions for the spectra between the observed bins, though this would in principle be possible given stellar spectra observed on shifted or irregularly-sampled grids.

We assume the spectral data ($\objdata_i$) have been observed with Gaussian noise that is uncorrelated between spectral bins, yielding a diagonal noise covariance matrix ($\objnoise_i$) for each star. Masked pixels are assigned unit flux and (effectively) infinite noise uncertainties. To account for the fact that the red clump might consist of multiple distinct sub-populations (or one population whose distribution of true spectra is non-Gaussian), we allow for multiple classes to exist in our model, each described by its own Gaussian process. We assume non-informative priors on the variables defining these Gaussian processes, adopting an infinite uniform prior on each mean and an inverse-Wishart prior on each class's covariance matrix \citep[p73]{Gelman_etal:2013}. We define the inverse-Wishart prior to have $\nb+1$ degrees of freedom, thereby placing a uniform prior on inter-pixel correlations, and a diagonal scale matrix ($\epsilon \, \identity$, with $\epsilon=10^{-6}$), minimizing the impact of the prior relative to the data. We infer the class membership of each star ($\objclass_i$) assuming they are sampled from categorical distributions with class probabilities ($\classprobs$) drawn, in turn, from a symmetric Dirichlet prior with concentration parameter $\alpha=1$.\footnote{The $n^{\rm th}$-order Dirichlet distribution is the set of $n$-dimensional lists with elements in the range 0 to 1 that sum to unity. It describes all ways to partition a dataset into $n$ classes, allowing for particular combinations to be preferred over others if desired.} These priors state our beliefs that, {\it a priori}: no location is preferred for the mean spectra; no scale is preferred for the covariance between two spectral bins; and the stars are as likely to be spread evenly between classes as they are to be concentrated in a single class. Our priors make no assumptions about (nor place any constraints on) the physics of the dataset, reflecting our desire for a purely data-driven inference. Should robust physical priors exist in another setting, it is simple to add them to the analysis.

\begin{figure}
	\includegraphics[width=\columnwidth]{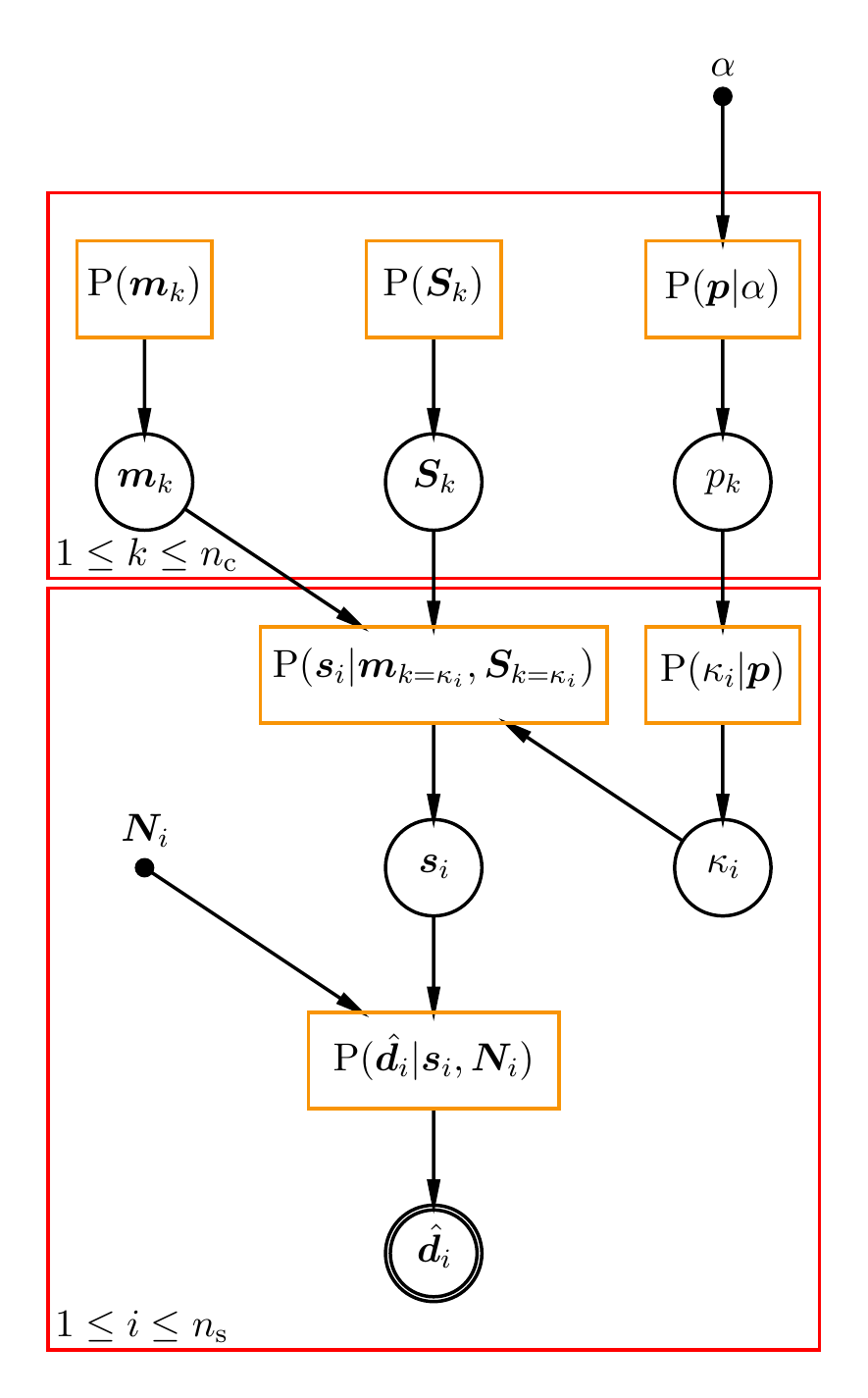}
    \caption{Network diagram for our hierarchical Bayesian model which is a graphical representation of our implemented modeling of the data. See Table 1 for the parameter descriptions.}
    \label{fig:network_diagram}
\end{figure}

The data, model parameters, priors and likelihood fully specify our probabilistic model of the APOGEE red clump dataset. This model is naturally hierarchical, with some parameters describing populations and others individual stars. This hierarchical nature is made clear in Figure~\ref{fig:network_diagram}, in which we plot the model as a network diagram. In this diagram, random variables are shown as single black circles, observables as double black circles and fixed parameters as solid black dots. Links between parameters are indicated by arrows, with the probabilistic relationships defining the links contained within orange boxes. The direction of these arrows indicates the order in which parameters must be drawn in order to forward-model the data. Finally, populations of objects are contained within red rectangles or plates, with the indices denoting membership of the population defined in the bottom left of the plate.

For clarity, we set out our model's parameters, data and constants in Table~\ref{tab:params} and the probability distributions defining each link in the top section of Table.~\ref{tab:prob_dists}. The particular set of probability distributions chosen allow for the conditional distributions of each model parameter to be written analytically: these conditional distributions are specified in the bottom section of Table.~\ref{tab:prob_dists}. We are therefore able to use Gibbs sampling~\citep{Geman_and_Geman:1984} to estimate the joint posterior. Gibbs sampling is a special case of Metropolis-Hastings Monte Carlo~\citep{Hastings:1970} in which a single iteration consists of redrawing each parameter in turn from its conditional distribution based on the current state of the sampler. For example, in our case, we first update the class probabilities, then the class memberships, the true spectra, and finally each class's mean spectrum and covariance matrix. Drawing proposed updates from the conditional distributions means the acceptance probability is, by definition, unity, yielding a highly efficient sampler even in  high-dimensional settings. By default, we initialize the sampler using the data, generating random class memberships before setting each class's mean spectrum and covariance matrix to the sample mean and covariance of the class members' data, and the true spectrum of each object to its observed data.\footnote{We obtain completely consistent results if we initialize the mean and true spectra to unity and the covariance matrix to the identity matrix.} The resulting sampler is written in Python and made publicly available on Github.\footnote{\href{https://github.com/sfeeney/ddspectra}{https://github.com/sfeeney/ddspectra}}

Our Gaussian Process model goes beyond standard approaches. As we sample the individual elements of the signal covariance matrix, the prior for the true spectra is very non-Gaussian. Were we to fit the covariance once and hold it fixed, as is common in the field, the true spectra would be Gaussian-distributed. By marginalizing over the covariance, however, we render these distributions very heavy-tailed, promoting sparse (i.e., typically small but occasionally large) excursions from the mean. As a result, we name the code SSSpaNG: {\em Stellar Spectra as Sparse, data-driven, Non-Gaussian processes.}

To demonstrate the effectively non-Gaussian nature of the prior on each star's true spectrum we can explicitly marginalize over the true signal covariance $\speccov$. Limiting ourselves to a single mixture component for clarity, we see that
\begin{align}
\prob\left(\objspec_i|\specmean\right) & = 
\int
	\prob\left(\objspec_i|\specmean,\speccov\right)
	\prob\left(\speccov\right)
d\speccov \nonumber \\
& \propto \int
	\left| \speccov \right|^{-\frac{\left( 2\nb+3 \right)}{2}}
	e^{-\frac{1}{2} {\rm Tr}\left( \left[ \left(\objspec_i-\specmean \right) \otimes \left(\objspec_i-\specmean \right) + \epsilon \identity \right]  \speccov^{-1} \right)}
d\speccov \nonumber \\
& \propto \left| \left(\objspec_i-\specmean \right) \otimes \left(\objspec_i-\specmean \right) + \epsilon \identity \right|^{-\frac{\left( \nb+2 \right)}{2}},
\end{align}
where the integral can be performed by identifying the integrand as an un-normalized inverse-Wishart distribution over $\speccov$. The result can be rewritten in the following suggestive form
\begin{equation}
\prob\left(\objspec_i|\specmean\right) \propto 
e^{-\frac{\nb + 2}{2} \ln{ \left| \left(\objspec_i-\specmean \right) \otimes \left(\objspec_i-\specmean \right) + \epsilon \, \identity \right|}},
\end{equation}
from which it becomes clear that this is a highly sparsifying prior whose negative logarithm is non-convex. Conceptually, it strongly prefers spectra close to the class mean, but if a spectrum differs greatly from the mean it is only penalized logarithmically. Note that the covariance prior's scale matrix, $\epsilon \, \identity$, acts to soften the prior, providing a small but non-zero floor to the determinant that reduces the preference for spectra exactly matching the mean. This reasoning explains why Gaussian-process modeling can outperform sparse image-reconstruction techniques~\citep{Sutter_etal:2014}.

\begin{table}
    \centering
    \caption{Model parameters, data and constants.}
    \label{tab:params}
    \begin{tabular}{ll}
        \hline
        quantity & description \\
        \hline
        $\ns$ & number of stars (29502) \\
        $\nc$ & number of classes (default: 1) \\
        $\nb$ & number of spectral bins (default: 343) \\
        $\specmean_k$ & mean spectrum of $k^{\rm th}$ class \\
        $\speccov_k$ & intrinsic spectral covariance of $k^{\rm th}$ class \\
        $\classprob_k$ & $k^{\rm th}$ class probability: fraction of stars in $k^{\rm th}$ class \\
        $\alpha$ & concentration parameter of Dirichlet prior on \\
         & class fractions \\
        $\objspec_i$ & true spectrum of $i^{\rm th}$ star \\
        $\objclass_i$ & class assignment of $i^{\rm th}$ star \\
        $\objdata_i$ & observed spectrum of $i^{\rm th}$ star \\
        $\objnoise_i$ & noise covariance matrix of $i^{\rm th}$ star \\
        \hline
    \end{tabular}
\end{table}

\begin{table*}
    \centering
    \caption{Priors, likelihoods and conditional distributions for Gibbs sampling. In our simplified notation, $\uniform$, $\dirichlet$, $\normal$ and $\invwish$ denote uniform, Dirichlet, normal and inverse-Wishart distributions, respectively.}
    \label{tab:prob_dists}
    \begin{tabular}{lll}
        \hline
        distribution & form & process \\
        \hline
        $\prob\left(\specmean_k\right)$ & $\uniform\left(-\infty,\infty\right)$ & Prior on $k^{\rm th}$ class's mean spectrum \\
        $\prob\left(\speccov_k\right)$ & $\invwish \left(\nb+1, \epsilon\,\identity \right)$ & Prior on $k^{\rm th}$ class's spectrum covariance \\
        $\prob\left(\classprobs|\alpha\right)$ & $\dirichlet\left(\alpha\right)$ & Prior on class probabilities \\
        $\prob\left(\objspec_i|\specmean,\speccov,\objclass_i\right)$ & $\normal\left(\specmean_{k=\objclass_i},\speccov_{k=\objclass_i}\right)$ & $i^{\rm th}$ object's spectrum as Gaussian Process \\
        $\prob\left(\objclass_i = k|\classprobs\right)$ & $\classprob_k$ & $i^{\rm th}$ object's class membership \\
        $\prob\left(\objdata_i|\objspec_i,\objnoise_i\right)$ & $\normal\left(\objspec_i,\objnoise_i\right)$ & Noisy, masked spectral measurements \\
        \hline
        $\prob\left(\specmean_k | \speccov_k, \objspec, \objclasses \right)$ & $\normal \left( \frac{1}{n_k} \sum_{\objclass_i = k} \objspec_i, \frac{1}{n_k} \speccov_k \right)$ & Conditional of $k^{\rm th}$ class's mean spectrum \\
        $\prob\left(\speccov_k | \specmean_k, \objspec, \objclasses \right)$ & $\invwish \left(n_k + \nb + 1, \scalemat_k + \epsilon \, \identity \right)$, & Conditional of $k^{\rm th}$ class's spectrum covariance \\
         & where  $\scalemat_k = \sum_{\objclass_i = k} \left( \objspec_i - \specmean_k \right) \otimes \left( \objspec_i - \specmean_k \right)$ & \\
        $\prob\left(\classprob_k | \objclasses, \alpha\right)$ & $\dirichlet\left(\alphas\right)$, where $a_k = \alpha + n_k$ & Conditional of class probabilities \\
        $\prob \left( \objspec_i | \specmean_{k=\objclass_i}, \speccov_{k=\objclass_i}, \objdata_i, \objnoise_i \right)$ & $\normal \left( \wfmean_i, \wfcov_i \right)$, & Conditional of $i^{\rm th}$ object's spectrum \\
         & where $\wfcov_i = \left( \speccov_{k=\objclass_i}^{-1} + \objnoise_i^{-1} \right)^{-1}$ &  \\
         & and $\wfmean_i = \wfcov_i \left( \speccov_{k=\objclass_i}^{-1} \specmean_{k=\objclass_i} + \objnoise_i^{-1} \objdata_i \right)$ &  \\
        $\prob \left( \objclass_i = k | \specmean, \speccov, \classprobs \right)$ & $ \frac{ \exp \left( -\frac{1}{2} \left[ \chi^2_{i,k} + \ln \left| \speccov_k \right| \right] + \ln \classprob_k \right) }{ \sum_{k^\prime} \exp \left( -\frac{1}{2} \left[ \chi^2_{i,k^\prime} + \ln \left| \speccov_{k^\prime} \right| \right] + \ln \classprob_{k^\prime} \right) }$, & Conditional of $i^{\rm th}$ object's class membership \\
         & where $\chi^2_{i,k} = \left( \objspec_i - \specmean_k \right)^T \speccov_k^{-1} \left( \objspec_i - \specmean_k \right) $ &  \\
        \hline
    \end{tabular}
\end{table*}



\section{Results}
\label{sec:results}

\subsection{Validation of Methodology: Predicting Unmeasured Spectral Regions}
\label{sec:validation}

To avoid the complications of comparing data gathered by different spectrographs, we validate our model and code by artificially masking a portion of one of our APOGEE spectra, namely the 15789 \AA\ cerium (Ce II) window of our lowest-SNR star (2M18335753-1302240), with an SNR measurement of SNR = 21, as listed in the APOGEE allStar file. We chose this feature, in particular, as it is a high-value detection in the APOGEE spectral region, being an s-process element. This feature was initially reported in \citet{Cunha2017}, who have provided measurements for a handful of the APOGEE stars. Measurements of this element for the full APOGEE survey would build on its chemodynamical reach. This would enable the mapping of the neutron capture family, in addition to the alpha, light and iron-peak elements, across the disk and into the halo and local group \citep[e.g.,][]{Majewski2017, Nidever2014, Hayden2015, Weinberg2019}. Nine windows were identified in \citet{Cunha2017}: we select one (unblended) Ce II window here (the line centered on 15784.75 \AA\ in air, converted to the vacuum scale of the APOGEE spectra) for validation of our methodology.

\begin{figure}
	\includegraphics[width=\columnwidth]{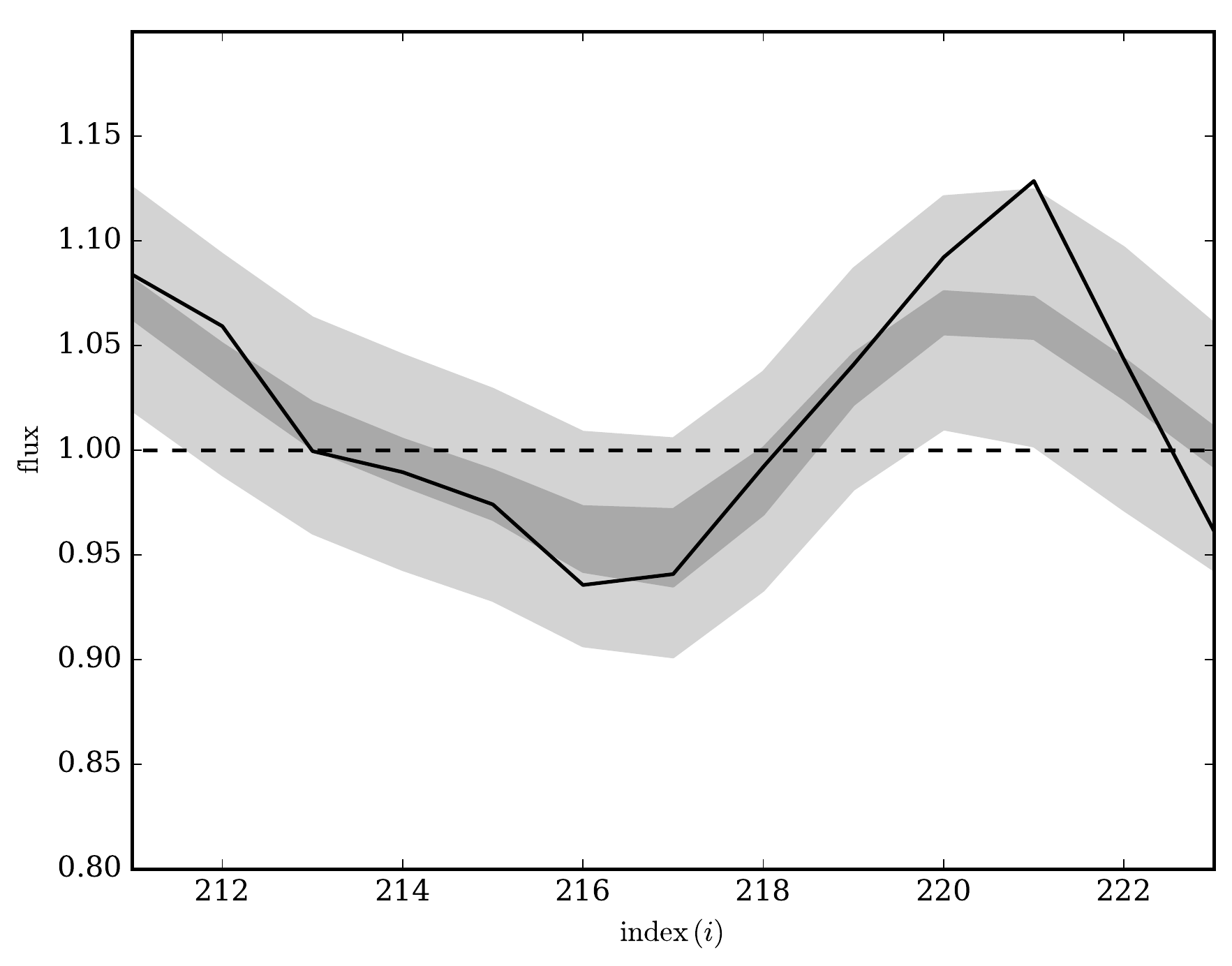}
    \caption{This Figure demonstrates the validation of our model and method via the recovery of an artificially masked portion of one star's spectrum: a 3 \AA\ region of spectrum centred on the cerium line at 15789 \AA\ (see Table 1). We select a star with a SNR of 21 for this demonstration to highlight the performance of the model for what would traditionally be considered very low SNR data. The measured spectrum in this region is shown as a solid black line; once masked (dashed line) the flux is set to one, with infinite uncertainty. After fitting our model with the APOGEE dataset (including the remainder of this star's measured spectrum) we find that the true spectrum for this star should most likely fall in the dark grey region, and the measured spectrum (i.e., including instrumental noise) should fall in the light grey region. This is in excellent agreement with the data.}
    \label{fig:recovery_test}
\end{figure}

The measured data for this star in the artificially masked region are plotted in Figure~\ref{fig:recovery_test} as a solid black line. The 68\% credible interval for the posterior probability on the star's true spectrum is plotted as dark grey, with the corresponding prediction for the observed spectrum (which also takes into account the [known] uncertainty on the observations) plotted in light grey. This prediction (strictly speaking, the posterior predictive distribution of the measured data) is in excellent agreement with the measured data, indicating that our model is capable of inpainting masked regions without bias. Note, in addition, that the uncertainty on the true spectrum is much smaller than the measurement noise, demonstrating our method's ability to denoise observed spectra by sharing information between stars (a phenomenon known as {\it shrinkage}~\citep[see, e.g.,][Chapter 13]{Busemeyer_etal:2015}).

This denoising property is relevant in the regime of extracting information from both weak lines and lower signal-to-noise data than typically required. In addition to the neutron capture element, Ce, the APOGEE spectral region has been shown to contain a number of r-process neodymium (Nd II) lines, which~\citet{Hasselquist_etal:2016} estimates are detectable in $\approx$ 18 percent of APOGEE spectra using equivalent width fitting techniques. Our expectation is this fraction will greatly improve given our Gaussian process modeling of the spectral lines, which can fit the true spectra of stars with lower uncertainty than the measurement noise.

We note that for this illustration we have inpainted one narrow window of a single star's spectrum, but this is generalizable to inpaint any spectral window, for any star. The predictive power to generate the spectra from the ensemble of all other stars and given prior measured spectral regions is detailed further in Sections~\ref{sec:inference} and~\ref{sec:info}.

\subsection{APOGEE Inference: Feature Correlations Across the Abundance Windows}
\label{sec:inference}

\begin{figure*}
	\includegraphics[width=2\columnwidth]{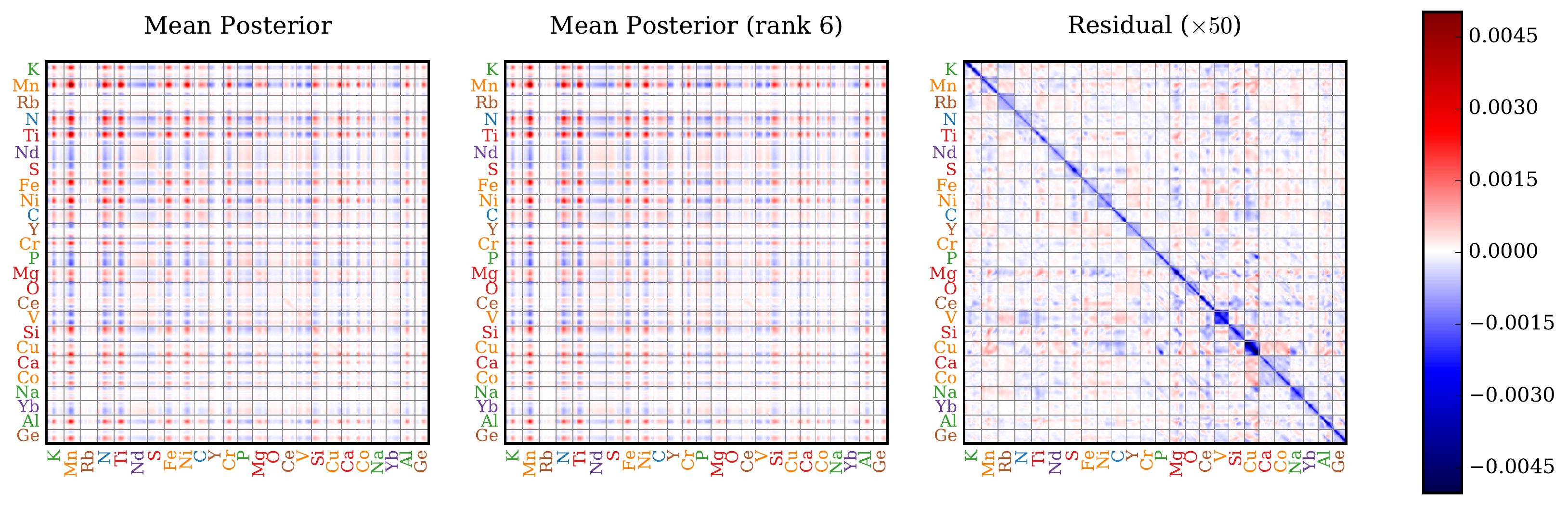}
    \caption{Left: the mean-posterior covariance matrix ($\speccov$) of the 343 spectral pixels that we model, with the corresponding colour bar giving the magnitude of this covariance (in units of flux$^2$). The divergent colour map shows the most positive and negative covariances in red and blue respectively and zero covariance as white. This matrix demonstrates that the spectral pixels are highly correlated. Centre: a reduced-rank approximation of the mean-posterior covariance matrix, constructed using only those eigenvectors with eigenvalues within $10^{-2}$ of the largest. This represents a factor of 57 reduction in the number of eigenvectors used to construct the mean-posterior covariance matrix. Right: the residual between the mean-posterior covariance matrix and its reduced-rank approximation, boosted by a factor of 50. This nearly diagonal residual shows that most of the variation between the denoised spectra is strongly correlated between spectra bins.}
    \label{fig:inferred_cov}
\end{figure*}

\begin{figure*}
	\includegraphics[width=2\columnwidth]{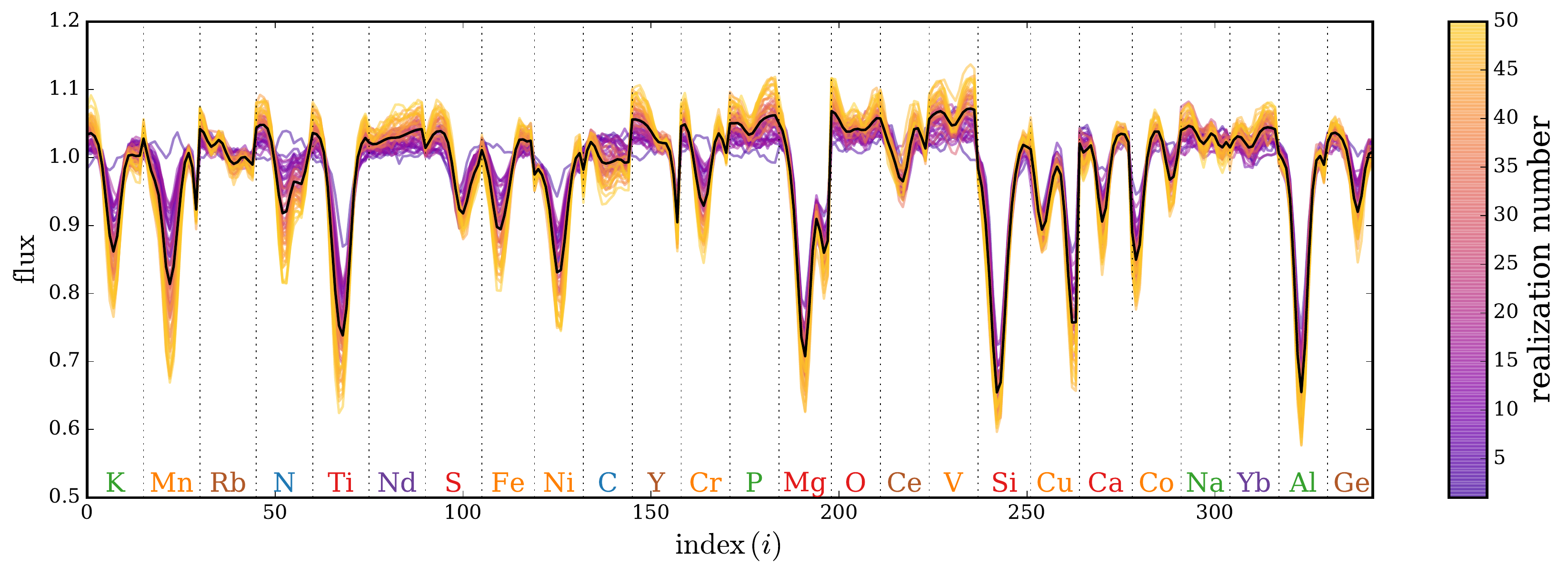}
    \caption{The mean-posterior mean spectrum ($\specmean$) of our Gaussian process model fit using the APOGEE data (black), along with 50 random realizations of potential true spectra ($\objspec$). These draws are coloured from purple to yellow according to their flux in the first spectral bin, and serve to demonstrate the correlations between pixels. Entirely uncorrelated data would show no structure in the colour gradient beyond the first bin; however, we see a clear stratification of yellow to purple as a function of the flux magnitude for most of the pixels.}
    \label{fig:gp_reals}
\end{figure*}

Our inference produces samples of the probability, mean spectrum and covariance matrix for each class considered, and the true spectrum and class membership of each object. Focusing initially on the single-class case, we plot our covariance and mean inference in Figures~\ref{fig:inferred_cov} and~\ref{fig:gp_reals}, respectively. We plot the mean-posterior covariance matrix in Figure~\ref{fig:inferred_cov} (left panel). The covariance has strong off-diagonal structure, indicating that certain spectral features are highly correlated and anti-correlated. Its eigenspectrum also decays rapidly: only 239 of 343 eigenmodes have eigenvalues larger than $10^{-4}$ of the maximum, and only six larger than $10^{-2}$ of the maximum. A low-rank approximation to the mean covariance retaining only the six largest eigenmodes is plotted in the centre panel of Figure~\ref{fig:inferred_cov}, and the resulting residuals (multiplied by a factor of 50 to render visible) in the right panel. Exploiting this decaying eigenspectrum by assuming the covariance is rank deficient would greatly reduce computation time (by a factor of roughly 187000 if six modes were retained!) but is left for future investigation.

The posterior mean of the mean spectrum is plotted in black in Figure~\ref{fig:gp_reals}. The mean spectrum is extremely well constrained: its 68\% credible interval is narrower than the width of the line. To illustrate the covariance structure captured by our model, we overlay 50 realizations drawn from our Gaussian process model conditioned on the APOGEE data, colour-coded by the value they take in the first spectral bin. These samples can be interpreted as examples of potential noiseless true spectra that could have led to the data. They illustrate the variability permitted by the model and highlight certain clear trends, most notably highly correlated differences in line depths.

\begin{figure*}
	\includegraphics[width=2\columnwidth]{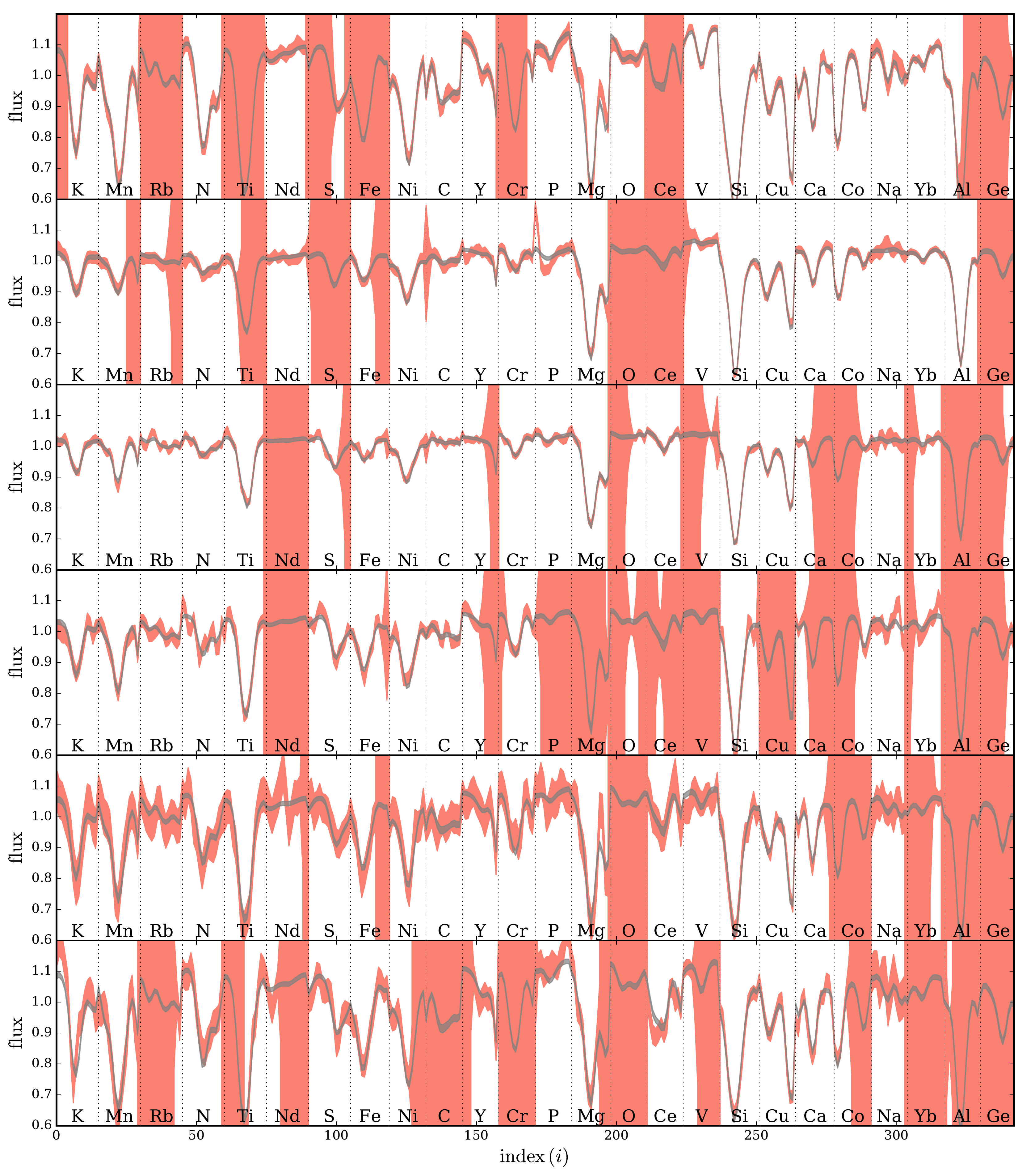}
    \caption{The measured and inferred spectra of six stars, all with low SNR (top to bottom: 49, 63, 75, 41, 21 and 23), selected to demonstrate our ability to both inpaint and denoise the data. The spectral regions shown are 3 \AA\ windows centred on the 25 elemental lines from Table~\ref{tab:window_centres}. The 68\% uncertainties on the observed spectra and inferred ``true'' spectra are shown as the pink and grey shaded regions, respectively (note the masked regions in the APOGEE spectra where the measurement uncertainties flare out to infinity). There is excellent agreement between the model and data. The first two spectra have completely masked cerium lines (15789 \AA), but our data-driven model makes a high-precision prediction of the cerium abundances for these stars". The middle two stars' neodymium (15372 \AA) lines are completely masked. Though the model again inpaints these regions successfully, the weakness of this line means recovery of significant neodymium abundances for these stars remains challenging. All other lines inferred by the model are denoised compared to the raw data, permitting higher precision  estimation of the  abundances.}
    \label{fig:inpainting_denoising_examples}
\end{figure*}

We demonstrate our inference of the true spectra of individual stars in Figure~\ref{fig:inpainting_denoising_examples}, selecting six illustrative examples. From top to bottom, we pick out two spectra whose 15789 \AA\ cerium windows are completely masked; two spectra whose 15372 \AA\ neodymium windows are fully masked; and the two lowest signal-to-noise spectra. The APOGEE IDs for these stars are 2M00014650+7009328, 2M00031631+0042234, 2M04480027+3337594, 2M06053121-0631412, 2M18335753-1302240 and 2M18295507-0340512, with signal-to-noise ratios of SNR = 49, 63, 75, 41, 21 and 23, respectively. Each panel of Figure~\ref{fig:inpainting_denoising_examples} contains two shaded regions. The pink shaded area indicates the 1-$\sigma$ deviations from the measured spectra due to noise (these are infinitely wide when the spectrum is masked); the grey, the 68\% posterior credible intervals on the true spectra.\footnote{Recall that we are inferring the true spectra at the measured spectral bins only. In this sense the smooth grey curves are perhaps misleading, as the posterior uncertainty is strictly infinite between data points.}

Figure~\ref{fig:inpainting_denoising_examples} clearly demonstrates our ability to inpaint masked regions of spectra and denoise low signal-to-noise spectra. The inpainting results for the cerium window are particularly encouraging. We are able to make precise (and very different) estimates of these two stars' spectra in the region of the cerium line, permitting, in principle, inference of their cerium abundances where none was previously possible. The same is true for, for example, for the aluminium lines of the third, fourth and fifth stars, along with the oxygen and germanium lines of the second, fifth and sixth stars. While we are also able to successfully inpaint the neodymium windows for the third and fourth stars, our model infers very weak line profiles in both cases, making an abundance inference challenging. Our ability to denoise the spectra is obvious for all stars considered: the uncertainties on the true spectra are in all cases smaller than the measurement uncertainty, permitting higher-precision abundance determination than previously possible. The Sodium line of the last two stars is a particularly good example of the potential for our method to denoise spectra.

The results presented up to this point assume that the APOGEE red clump stars belong to a single class (and their true spectra are therefore realizations of a single Gaussian process). We have experimented with allowing multiple classes, initializing the sampler with random class memberships; however, we find little impact on our final results. The sampler finds slight differences between the classes' mean spectra ($\specmean_k$) and covariances ($\speccov_k$), but these are driven by the initial randomized class memberships: very few stars leave one class for another during the sampling process, and those that do typically do so only once, in the sampler's first iteration. This is because the probability distribution used for drawing a star's class membership (Table~\ref{tab:prob_dists}, last row) drops exponentially with the squared distance between the star's true spectrum and each class's mean spectrum\footnote{Specifically, the Mahalanobis distance, or number of ``sigma'' the star's spectrum is from the class's mean.}. In very high dimensions, for almost all stars the distance to a new class is typically much larger than the distance to the current class, and the probability of transitioning to a new class is essentially zero. As such, we believe the class assignments are strongly dependent on the choice of initial state of the Markov chain and hence not meaningful. Exploring these high-dimensional clustering issues is left to future work. 

\subsection{The Measured Information Content in the Spectra}
\label{sec:info}

\begin{figure*}
	\includegraphics[width=\columnwidth]{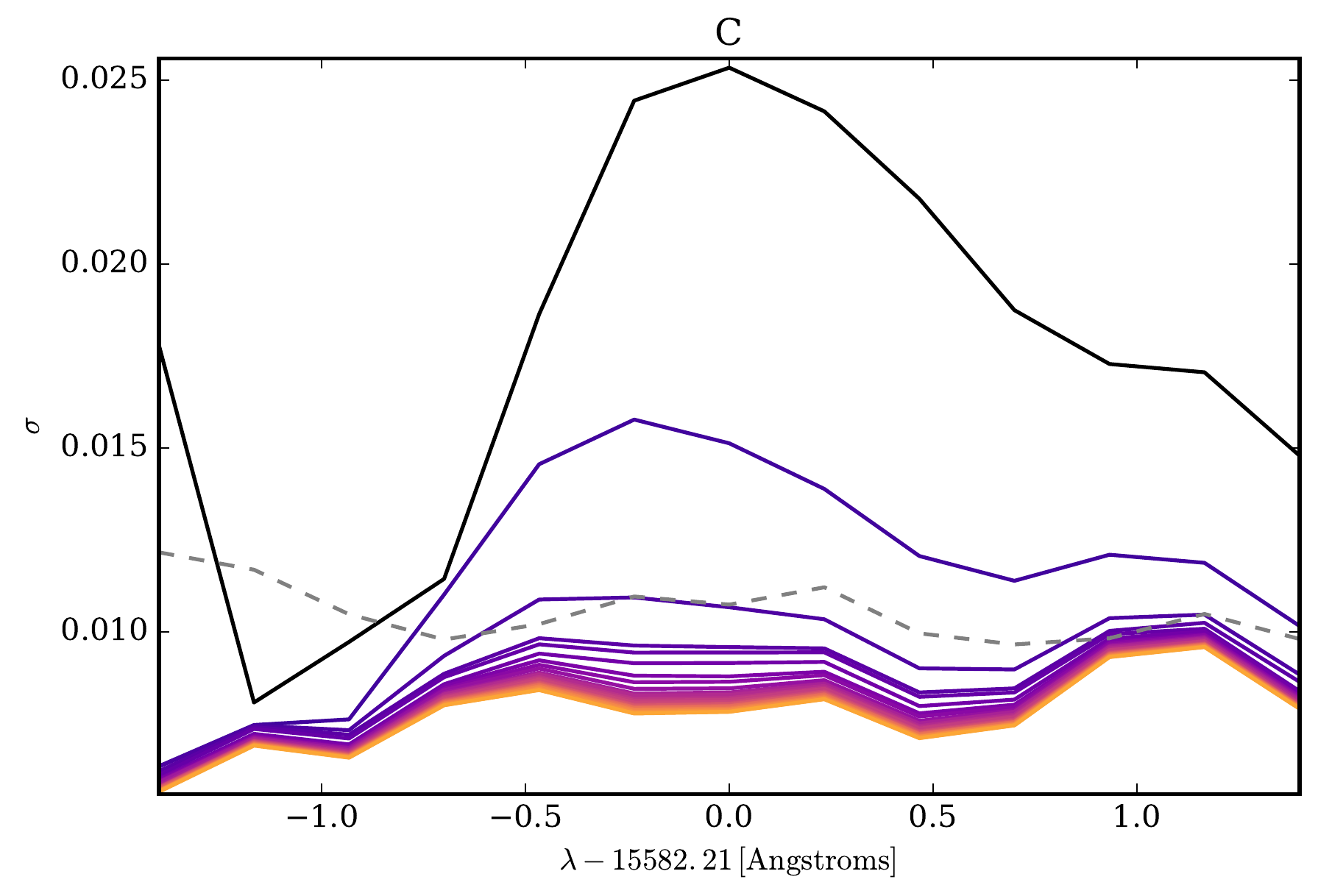}
	\includegraphics[width=\columnwidth]{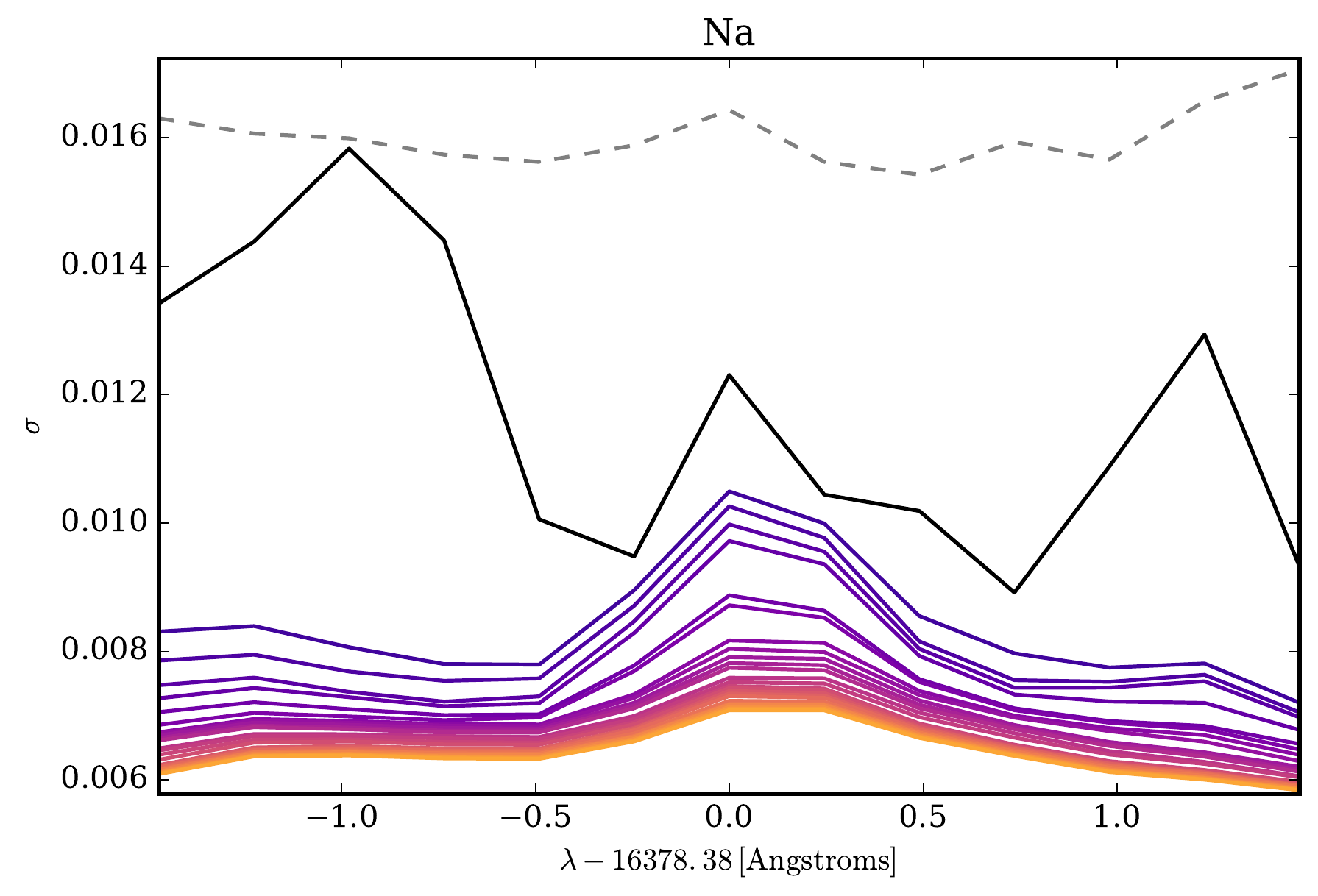}
	\includegraphics[width=\columnwidth]{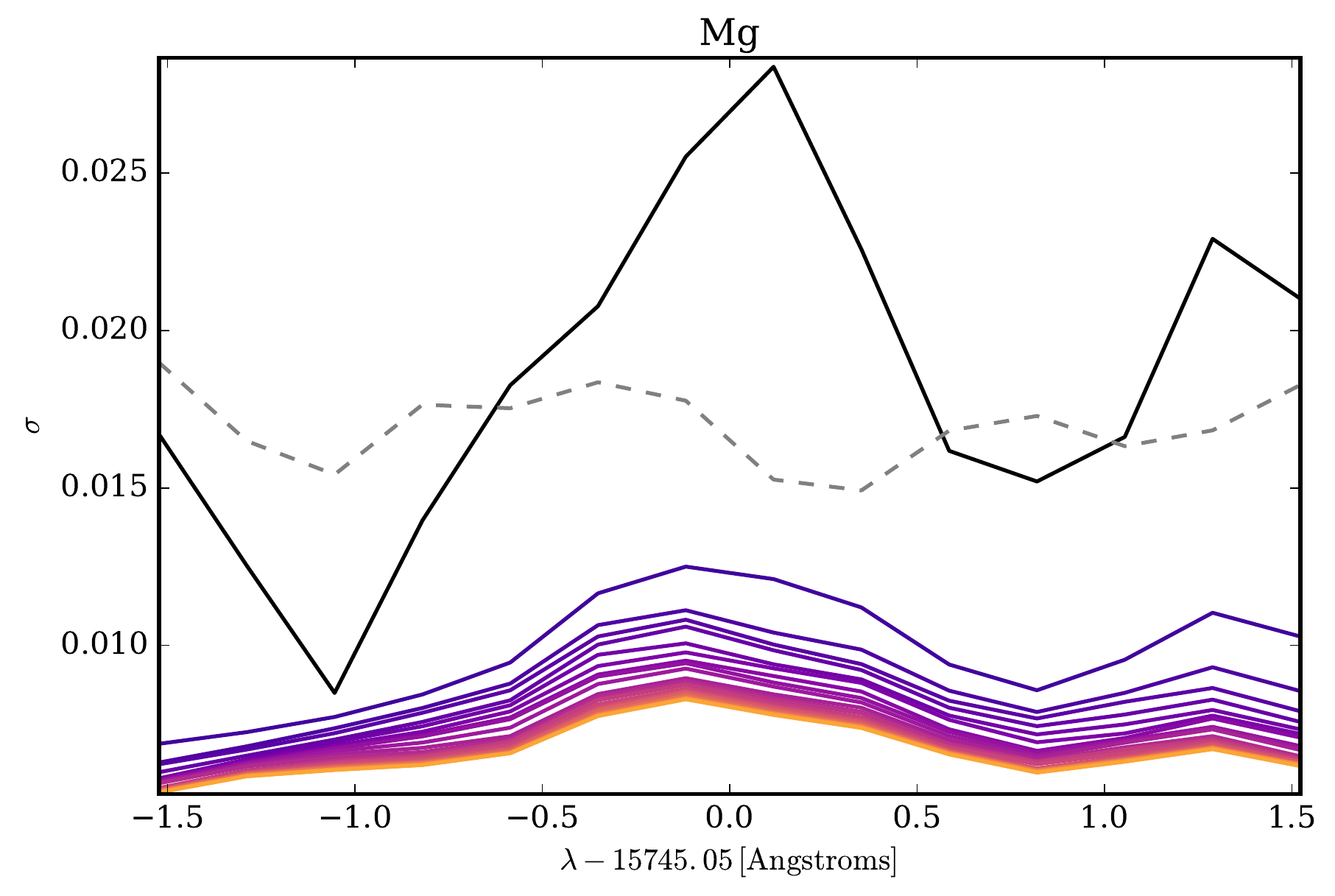}
	\includegraphics[width=\columnwidth]{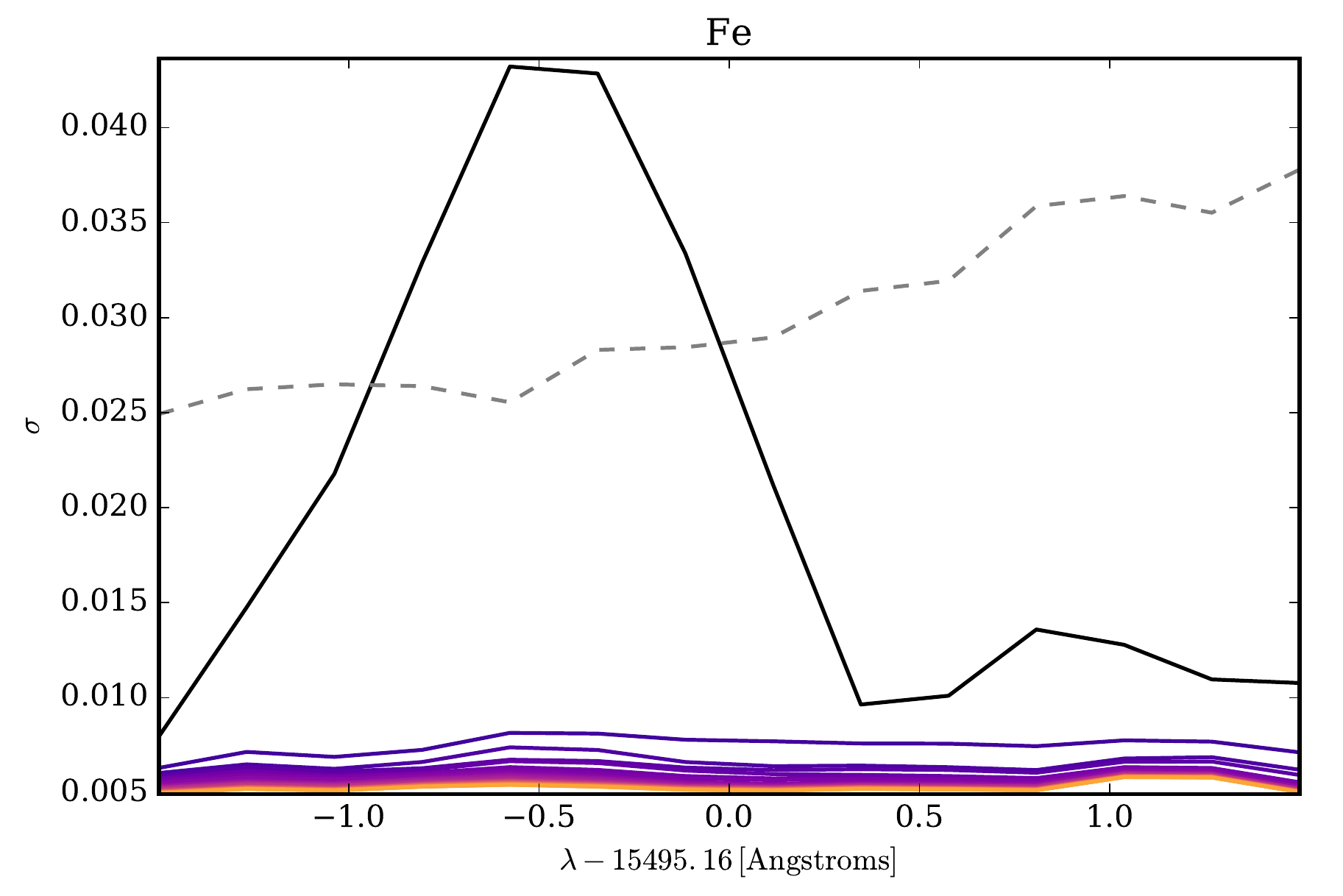}
	\includegraphics[width=\columnwidth]{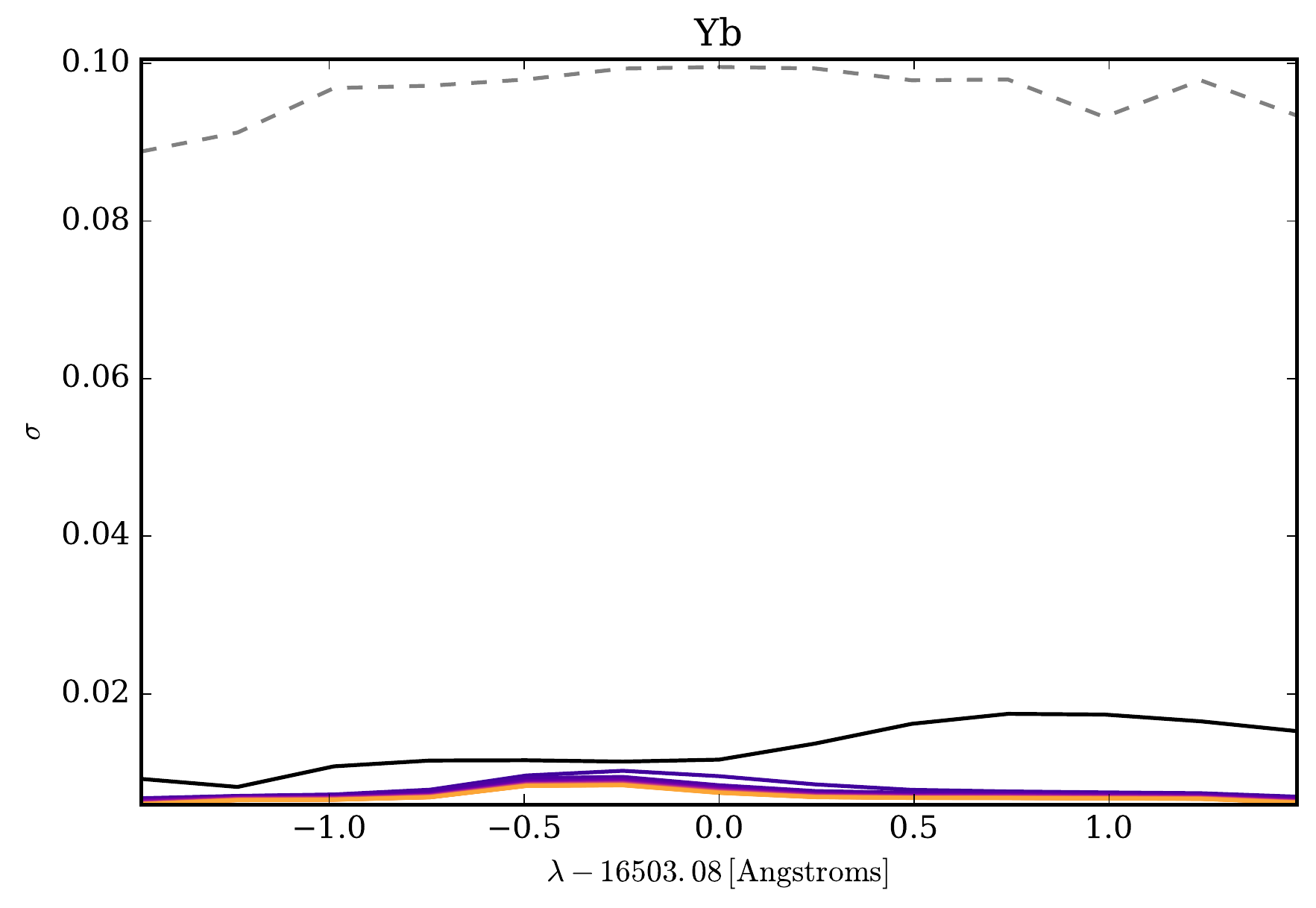}
	\includegraphics[width=\columnwidth]{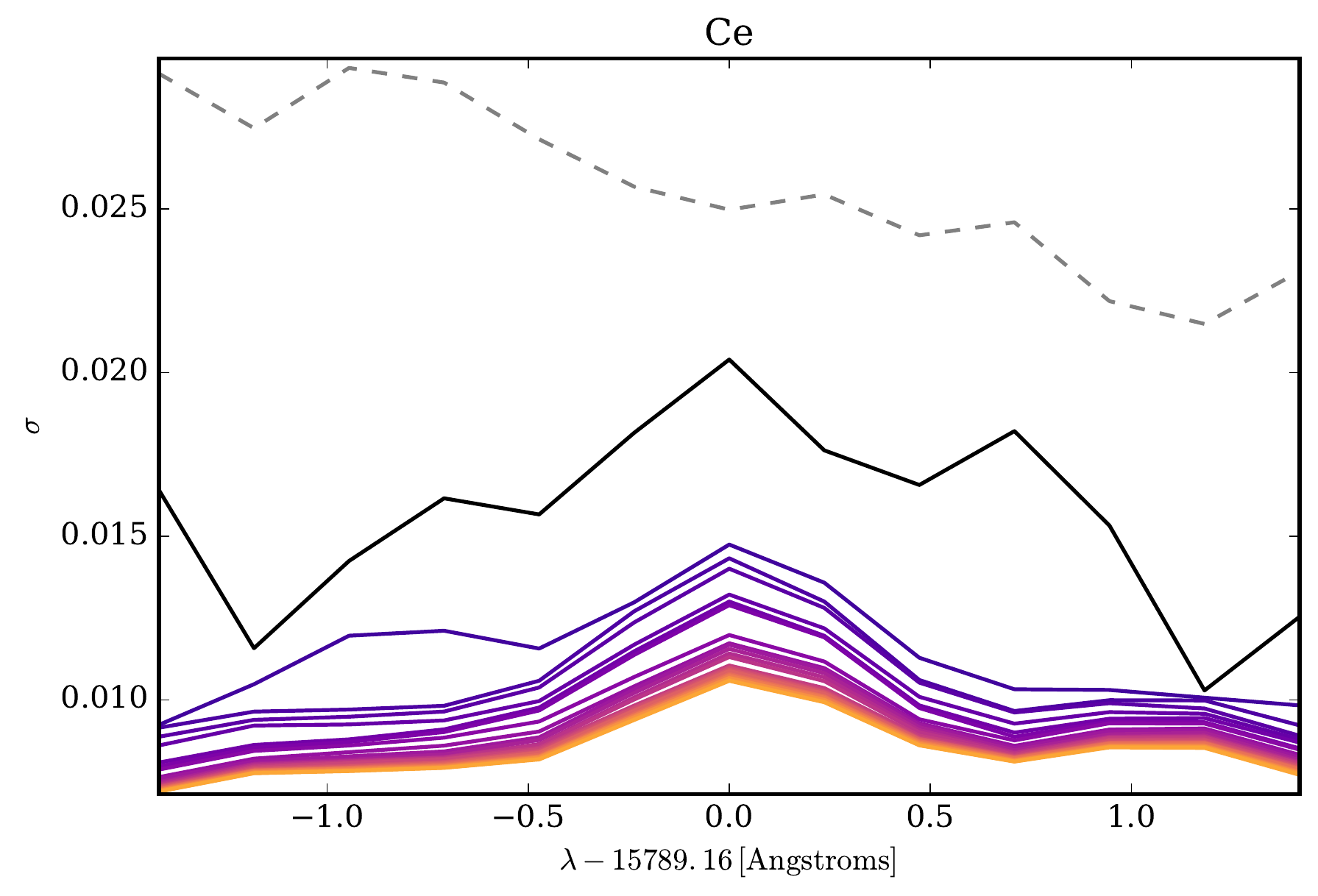}
    \caption{Root-mean-square uncertainties on the spectra within our illustrative set of elemental windows, centred on features due to C, Na, Mg, Fe, Yb and Ce, respectively. The black line shows the uncertainty on the predicted spectrum of a new APOGEE star having not observed any portion of its spectrum; the grey dashed line indicates the typical uncertainties due to APOGEE noise. The remaining lines show how the uncertainty decreases after having perfectly observed the $1 \le n \le 24$ most informative elemental windows of the new star's spectrum, coloured from purple (most informative) to yellow (least informative). The order in which elements are added is plotted in Figure~\ref{fig:single_element_information}. Note that the impact of adding observations decreases as the information gain curves of Figure~\ref{fig:single_element_information} become less steep.}
    \label{fig:single_element_errs}
\end{figure*}

We now turn to quantifying the information contained in each elemental window. Our aim is to determine the regions of spectra that are most informative about particular elements of interest. We must note, however, that our elemental windows can contain spectral features in addition to the central absorption line, and thus strong correlations between two windows are not necessarily due solely to the central elements themselves. We start with the mean-posterior covariance within each window, $\bar{\speccov}_{XX}$, as this describes the fundamental uncertainty with which we can predict the true spectrum of a new red clump star having observed our APOGEE sample. The subscript $X$ here denotes the spectral bins defining the elemental window of interest. We summarize this covariance matrix for six elemental windows ($X=\{{\rm C, Na, Mg, Fe, Yb, Ce}\}$: one from each elemental family) by plotting the root-mean-square (RMS) uncertainty,
\begin{equation}
\bm{\sigma}_X = \sqrt{{\rm diag} \left[\bar{\speccov}_{XX}\right]},
\end{equation}
in black in Figure~\ref{fig:single_element_errs}. For context, we overlay the typical measurement uncertainty\footnote{The square root of the average noise variance in each spectral bin, where the average is taken over stars in whose spectra the bin is not masked.} as a grey dashed line. This immediately demonstrates that our model of the APOGEE spectra allows us to make sub-noise predictions for some portions of a new star's spectrum without taking further data. The results for the ytterbium window are especially interesting, as the average instrumental noise seems particularly large in this region.

To determine which windows are the most informative, imagine observing one window of this new star's spectrum (corresponding to, say, element $Y$) {\it without measurement error}. The long-range correlations present in the inferred covariance matrix (i.e., the fact that elemental abundances are determined by a finite number of physical processes) imply that by doing so we should better constrain the elemental window of interest. To quantify the information gained about element X by (perfectly) observing element Y, we calculate the conditional covariance matrix
\begin{equation}
\condcov_{XX|Y} = \bar{\speccov}_{XX} - \bar{\speccov}_{XY} \bar{\speccov}_{YY}^{-1} \bar{\speccov}_{YX}.
\end{equation}
The conditional covariance contains our full prediction for the uncertainty on window $X$ having observed window $Y$, but we must compress it in order to construct a useful metric for quantifying information gain. We therefore define our information gain metric to be
\begin{equation}
I = \frac{1}{2} \log \frac{ \left| \speccov_{XX} \right| }{ \left| \condcov_{XX|Y} \right| } \ge 0.
\end{equation}
This can be interpreted in two ways. From an information theory perspective, the differential entropy of an $n$-dimensional multivariate normal distribution with covariance $\speccov_{XX}$ is $\frac{n}{2} \left[1 + \log 2\pi \right] + \frac{1}{2} \log |\speccov_{XX}|$~\citep[see, e.g.,][Chapter 9]{Cover_Thomas:2006}. Changing the covariance matrix to $\condcov_{XX|Y}$ as we do by observing additional windows therefore changes the differential entropy of the system (i.e., adds information to it) by precisely $I$ nats. From a geometric perspective, note that the determinant of a matrix is the hypervolume of the ellipsoid whose major axes are the eigenvectors of the matrix and have length of the eigenvalues. The square root of the determinant of a {\it covariance} matrix is therefore the volume of the error ellipsoid on the quantities of interest, up to a constant prefactor. Our metric $I$ can therefore also be interpreted as the logarithmic factor of improvement in predicting window $X$'s true spectrum obtained by observing window $Y$. Regardless of the interpretation, observing a new window can only add information, contracting the covariance matrix (or, in the worst-case scenario, leaving it unchanged), and thus $I$ cannot be less than zero.

With this metric in hand, we can take each elemental window in turn and determine the information gained by observing each other window. The window $Y^1$ with the most negative $I$ is the most informative about our target window $X$; indeed, as our metric $I$ is symmetric, these two elements are the most informative about each other. We then repeat this process, conditioning on $Y^1$ {\it and} each other window in order to find the second most informative window, $Y^2$, continuing to add windows until we find the optimal order in which to build up information on the element of interest. We denote the list of the $n$ most informative elements $\bm{Y}^n = \{ Y^1, Y^2, \ldots, Y^n \}$; the covariance in window $X$ conditioned on these elements is $\condcov_{XX|\bm{Y}^n}$.

\begin{figure*}
	\includegraphics[width=\columnwidth]{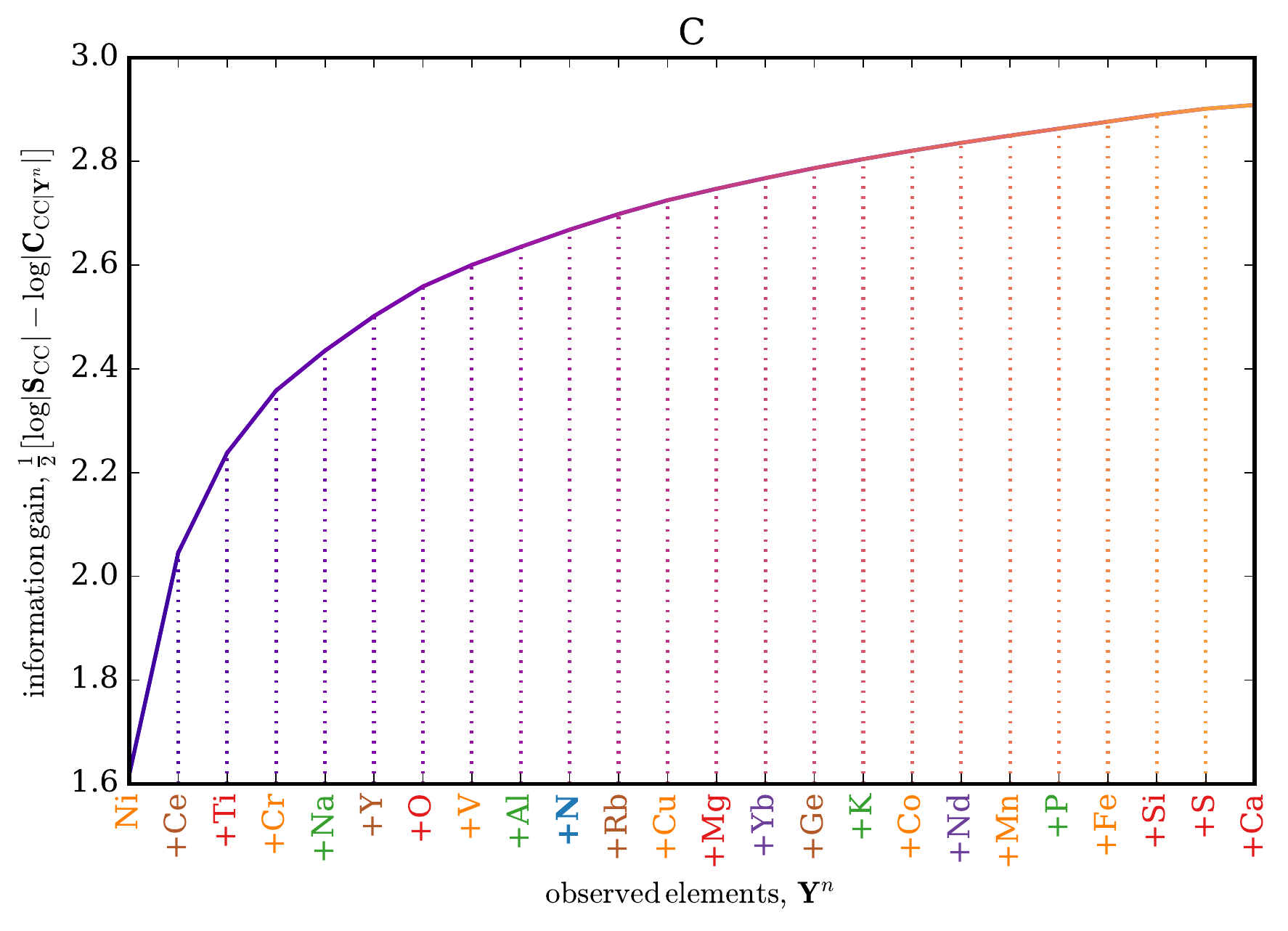}
	\includegraphics[width=\columnwidth]{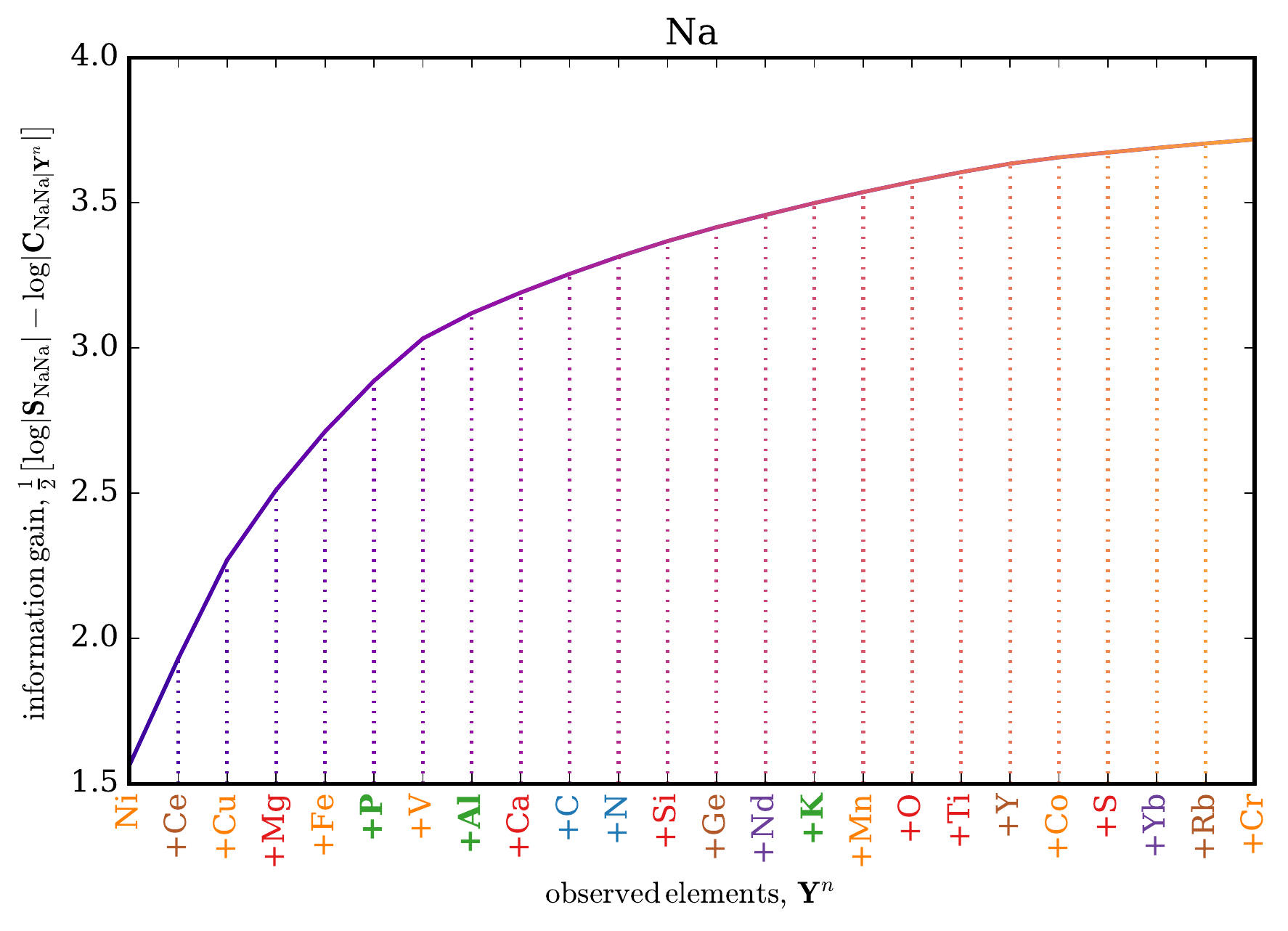}
	\includegraphics[width=\columnwidth]{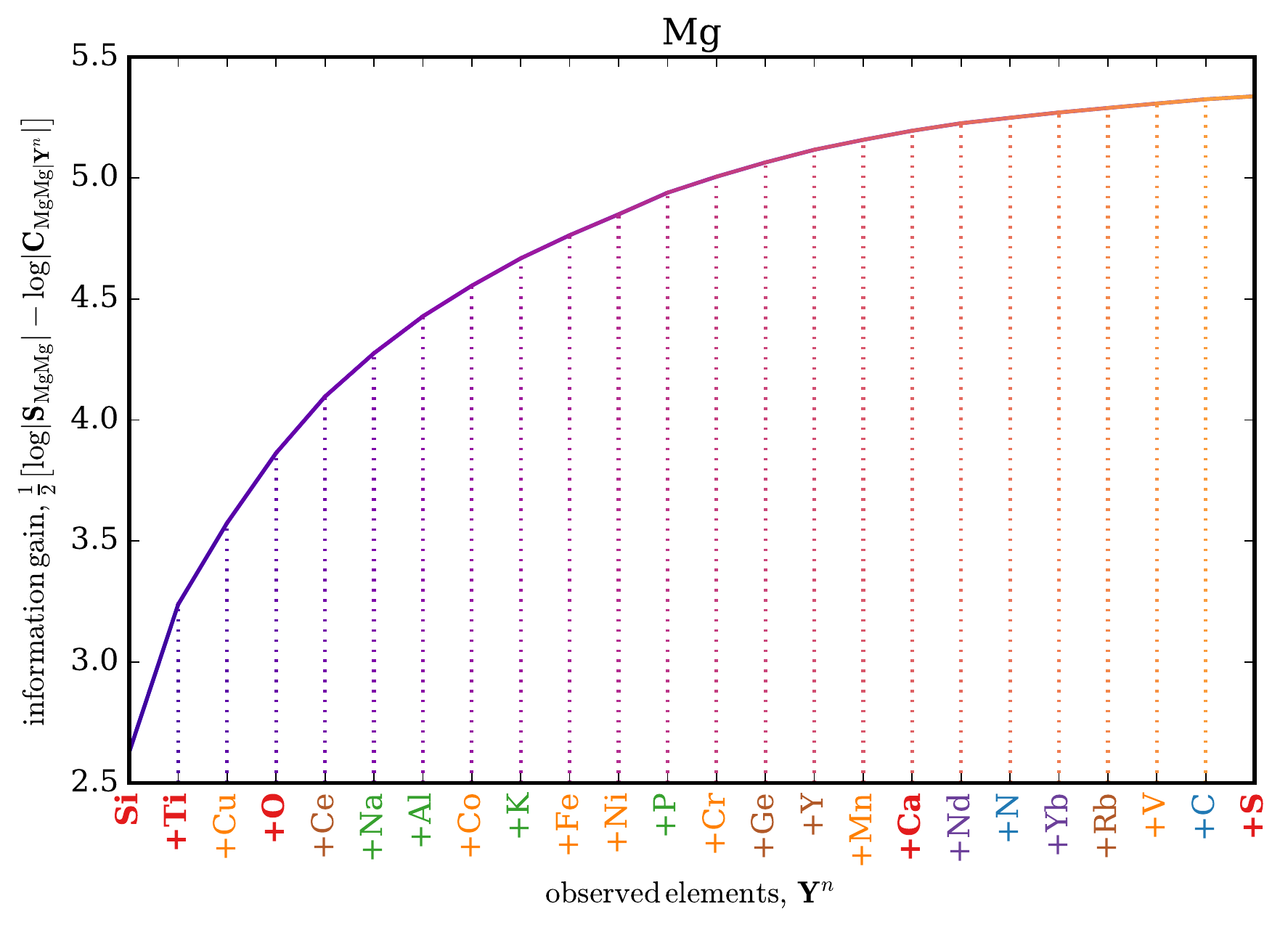}
	\includegraphics[width=\columnwidth]{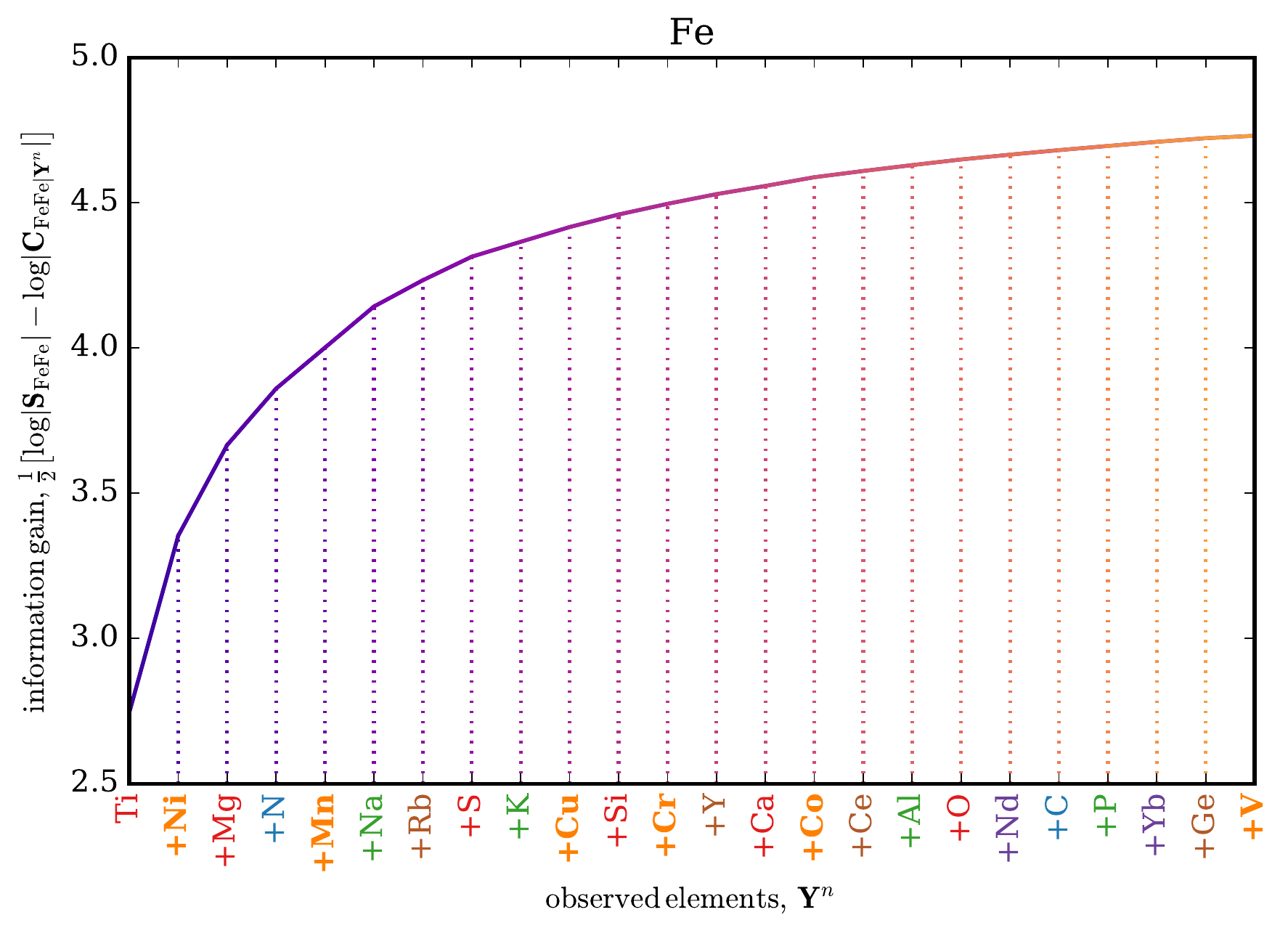}
	\includegraphics[width=\columnwidth]{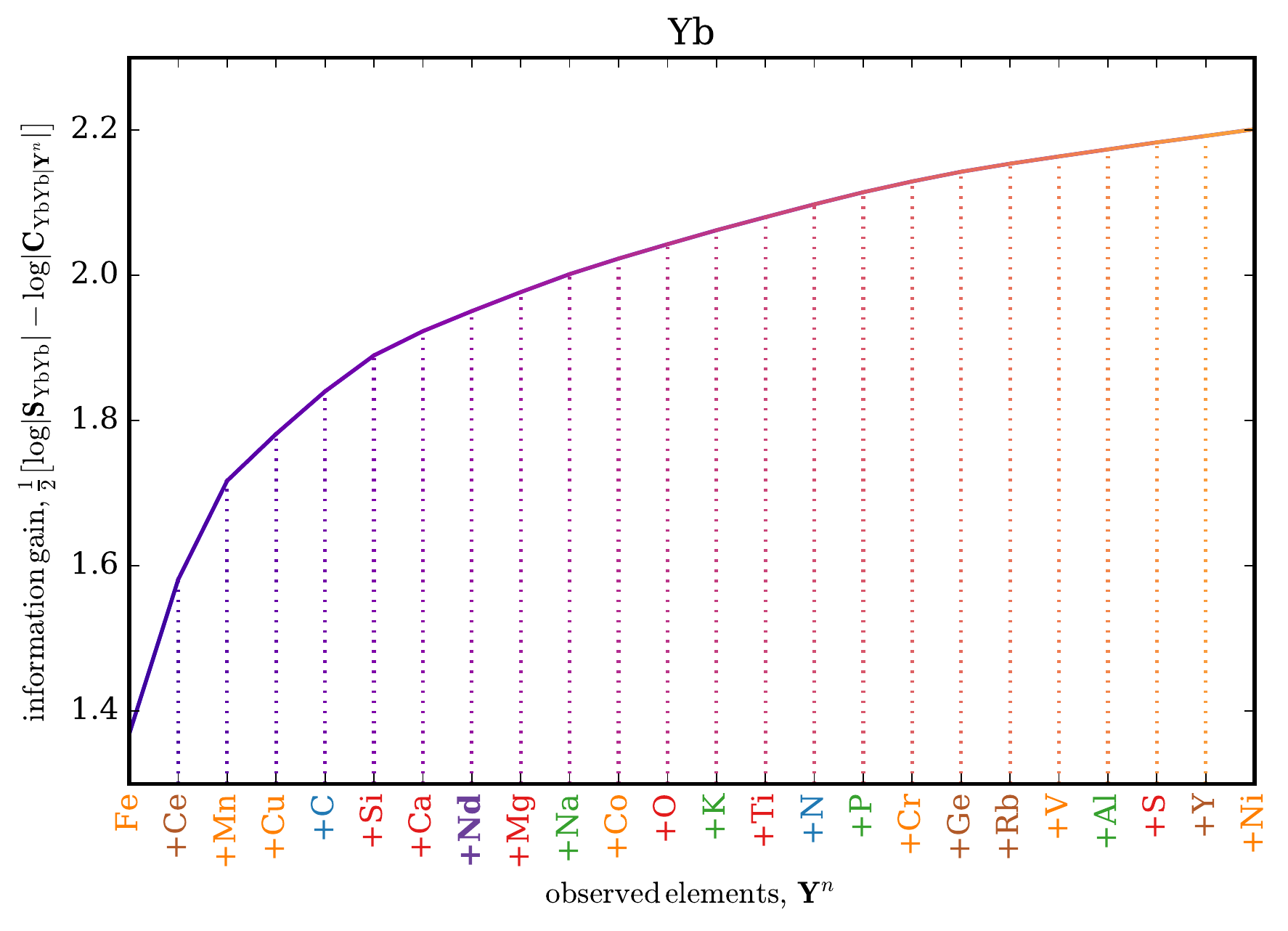}
	\includegraphics[width=\columnwidth]{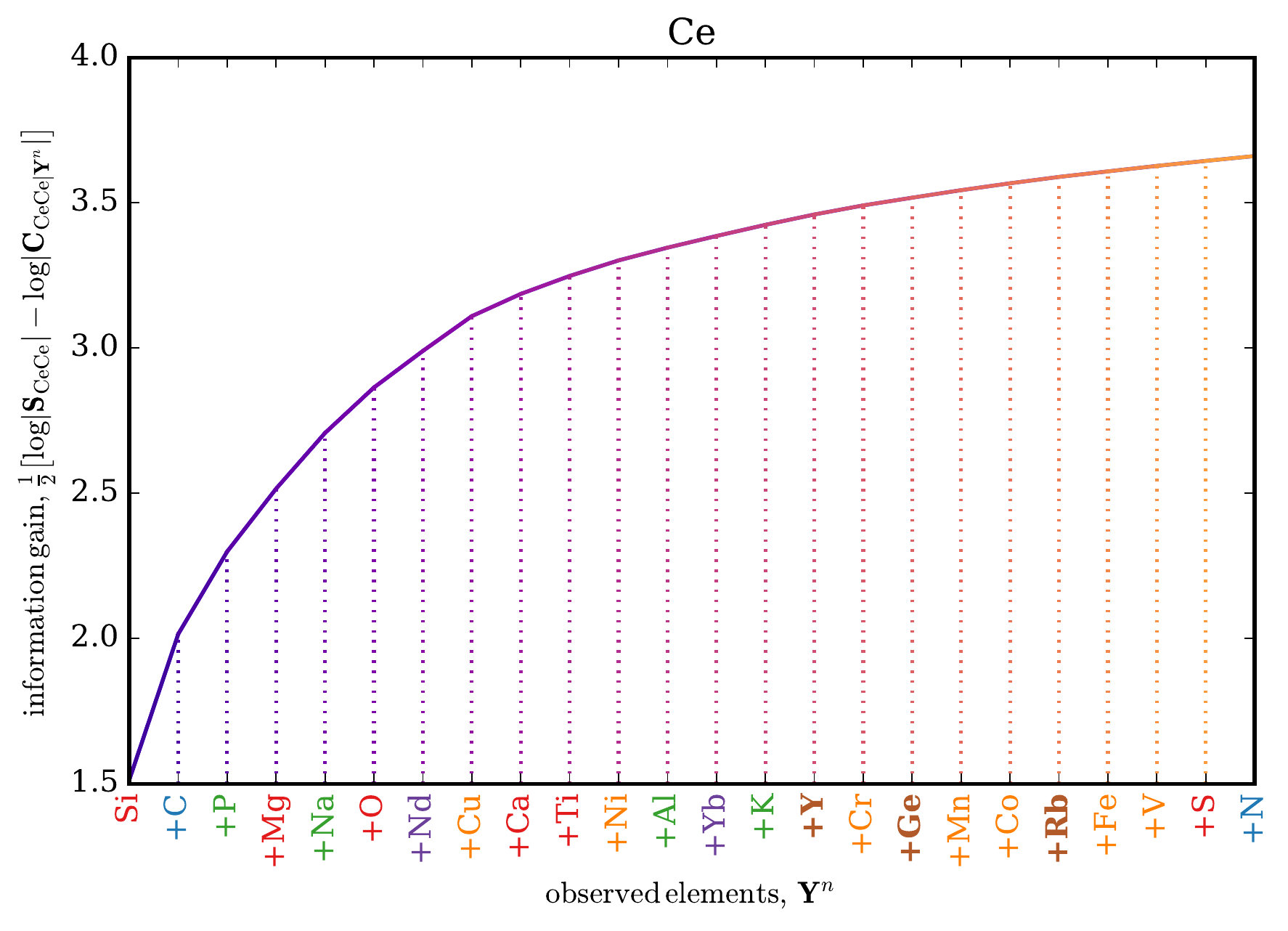}
    \caption{Information gains for our illustrative elemental windows obtained by observing the other 24 windows. The x-axis of each panel lists, from left to right, the window that would provide the most information on the element of interest assuming all previous windows have been observed. To take the top left-hand panel as an example: one would learn the most about the C window by observing Ni, then adding Ce, Ti, Cr {\it et cetera}. The y-axis quantifies the resulting information gain, and can be interpreted as a change in entropy of the system or the factor of reduction in the total uncertainty on the target window's predicted spectrum provided by observing the other windows. Note the different y-axis ranges for the six different elements (the most extreme being Yb and Mg): the larger the overall information gain, the better the elemental window is predicted by the rest of the spectrum. Note also that while the gain from observing successive elements decreases it does not entirely flatten: each individual element adds information on the target element. Finally, the finite range of these plots indicates that, though elements are highly correlated, no one perfectly predicts another.}
    \label{fig:single_element_information}
\end{figure*}

We plot the results of this process for the six illustrative elements in Figures~\ref{fig:single_element_errs} and~\ref{fig:single_element_information}. In Figure~\ref{fig:single_element_errs} we demonstrate how the RMS uncertainty within each elemental window shrinks as we condition on more and more information, now taking the RMS uncertainty to be
\begin{equation}
\bm{\sigma}_X = \sqrt{{\rm diag} \left[\condcov_{XX|\bm{Y}^n}\right]}.
\end{equation}
We plot the RMS uncertainties after conditioning on the $1 \le n \le \nb-1$ most informative windows as a series of solid curves, coloured from purple to yellow. Observing the most informative window, $Y^1$, significantly improves the uncertainty on the spectral window of interest, and conditioning on additional windows continues to add information, albeit with diminishing returns. After observing all other windows, the RMS uncertainty at the centre of the window of interest (i.e., directly over the elemental absorption line) has been reduced by a factor of roughly two to five.

In Figure~\ref{fig:single_element_information} we plot the most informative windows for our six elements of interest first, along with the information gained by observing each additional window moving to the right on the x-axis. The windows' labels are coloured by their elemental family, with members of the target element's family picked out in bold. Recall that our information gain metric can be interpreted as the logarithm of the fractional reduction in volume of the error ellipses on the true spectrum. These plots cover the rough range $1.4 \le I \le 5.3$, corresponding to reducing the error volume by factors of 4 to 200. Reflecting the qualitative results of Figure~\ref{fig:single_element_errs}, each of the curves in Figure~\ref{fig:single_element_information} flattens as more elements are observed, indicating that the single greatest information gain is provided by observing the most informative elemental window and the bulk of the information is provided by the first 10 or so elements. None of the curves plateau, however, and thus all elements provide information on the window of interest. It is perhaps interesting to note that the most informative element is not, in general, from the same family as the element of interest (though this is true for magnesium). We caution over-interpretation of this point, however, for two reasons: 1) this conclusion applies only to this specific set of spectral windows; and 2) these windows are broader than the elemental features they are designed to capture, and can therefore contain information about a number of elements.

\begin{figure*}
	\includegraphics[width=\columnwidth]{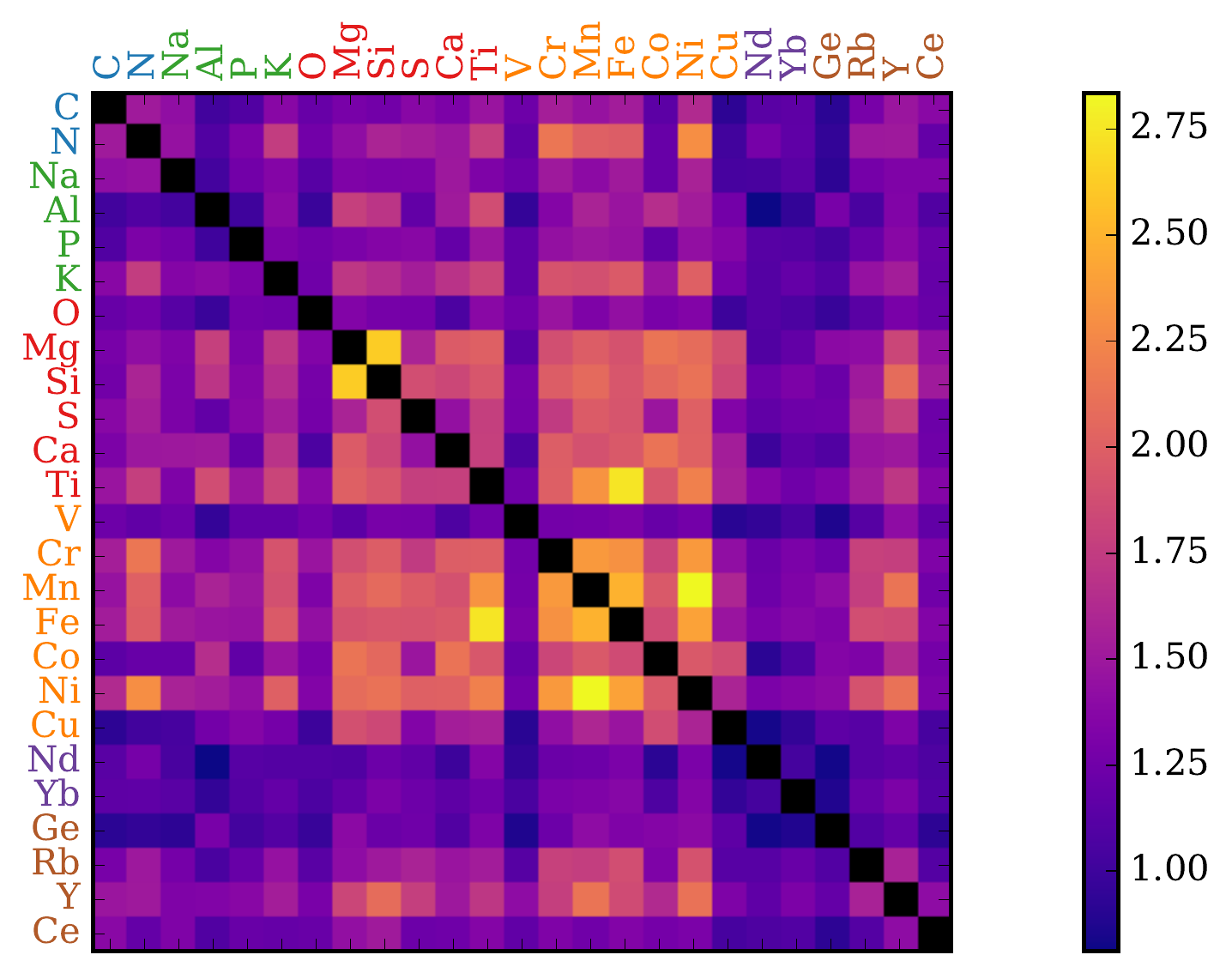}
	\includegraphics[width=\columnwidth]{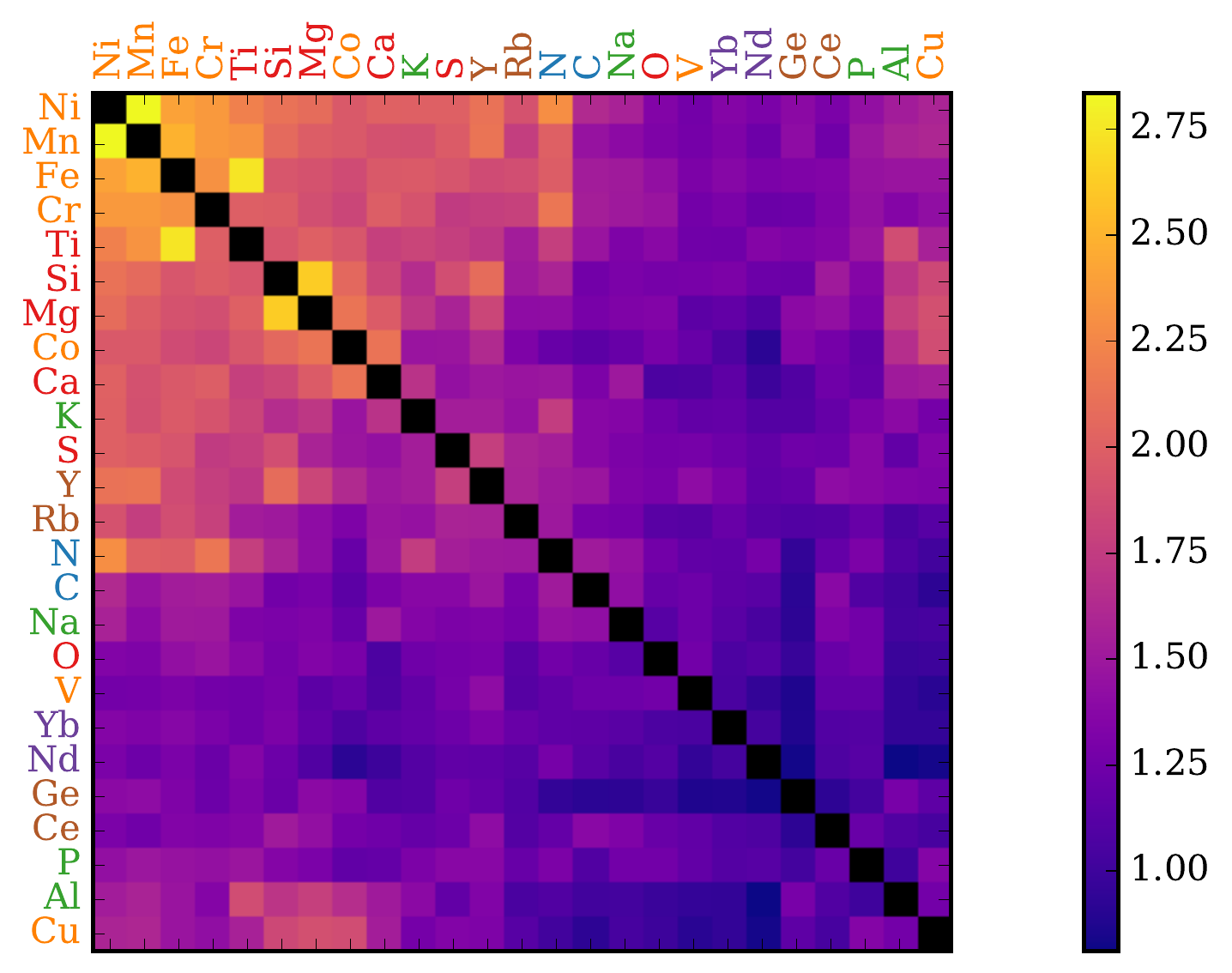}
    \caption{Information gains for pairs of elemental windows, colour coded from purple to yellow in order of increasing information gain. In the left panel the elemental windows are grouped according to their nucleosynthetic family, as indicated by the colour of their label. The iron-peak family of elements (and Ni, Mn, Fe and Cr in particular) are the most predictive, followed by the alpha elements (Ti, Si and Mg in particular). In the right panel, the windows are sorted to minimize the difference between adjacent rows, thereby clustering elemental windows with similar information content. Note that this does not discretely separate elements into their nucleosynthetic families, particularly beyond the iron-peak and alpha elements.}
    \label{fig:all_element_information}
\end{figure*}

Having discussed our detailed findings for the six illustrative elemental windows, we now summarize the results for all of the elemental windows. In Figure~\ref{fig:all_element_information} we plot the information gains for every pair of windows; that is, for each elemental window we plot the information we would gain by observing each other window perfectly. As we have demonstrated in Figures~\ref{fig:single_element_errs} and~\ref{fig:single_element_information}, there is much information to be gained by adding further observations, but given there are 24! ways of ordering them we will have to make do with the first. In doing so, we at least discover the most informative elemental pairs. We present the complete set of information gains in two ways. In the left panel of Figure~\ref{fig:all_element_information}, we group the elements by their families, sorting within each family by increasing atomic number. The most informative elemental pairs (the brightest yellow pixels) are Ni-Mn (both iron-peak), Mg-Si (both alpha) and Fe-Ti (iron-peak and alpha), and this trend is generically true of the families as a whole: the iron-peak and alpha elements predict both themselves and each other well. Indeed, these elements also predict the other families well.

There is considerable structure in this matrix, with patterns of predictivity common to multiple elements: for example, the majority of alpha-element and iron-peak rows look very similar. We make a first pass at sorting using this structure in the right panel of Figure~\ref{fig:all_element_information}. We quantify the similarity between the $i^{\rm th}$ and $j^{\rm th}$ rows in the plotted matrix of information gains using the distance
\begin{equation}
d(i \leftrightarrow j) = \sum_k |I_{ik} - I_{jk}|,
\end{equation}
where $I_{ik}$ is the information gain for the $i^{\rm th}$ element from observing the $k^{\rm th}$.\footnote{Note that we use an absolute distance metric here: using a Euclidean distance metric instead yields similar results.} To sort the elements by similar predictivity we use a simple greedy algorithm, approximating the global optimum through a series of locally optimal choices. To start, we pick an initial value of $i$, then find the most similar element by determining the row $j$ that minimizes $d(i \leftrightarrow j)$. We then take element $j$ as the comparator, calculating distances ($d(j \leftrightarrow k)$) to find the most similar of the remaining elements, and repeat until no elements remain. This approach is not guaranteed to find the global optimum, and indeed depends on the first element chosen. We therefore repeat the process with each element as the starting point and select the sorted matrix whose total distance between rows is minimal.

The resulting sorted matrix, plotted in the right panel of Figure~\ref{fig:all_element_information}, has much clearer structure than when sorted by elemental family. On the whole, the iron-peak elements are most similar as well as most predictive, closely followed by the alpha elements; copper, vanadium and oxygen are, however, notable exceptions to these patterns. There is also a fairly clean break around rubidium and nitrogen, beyond which the information gains drop noticeably. Note, however, that aluminium and copper are moderately informative about titanium, silicon, magnesium and cobalt. Our ordering placed them beyond the Rb-N break; this may well be due to the sub-optimality of the greedy algorithm.

\begin{figure}
	\includegraphics[width=\columnwidth]{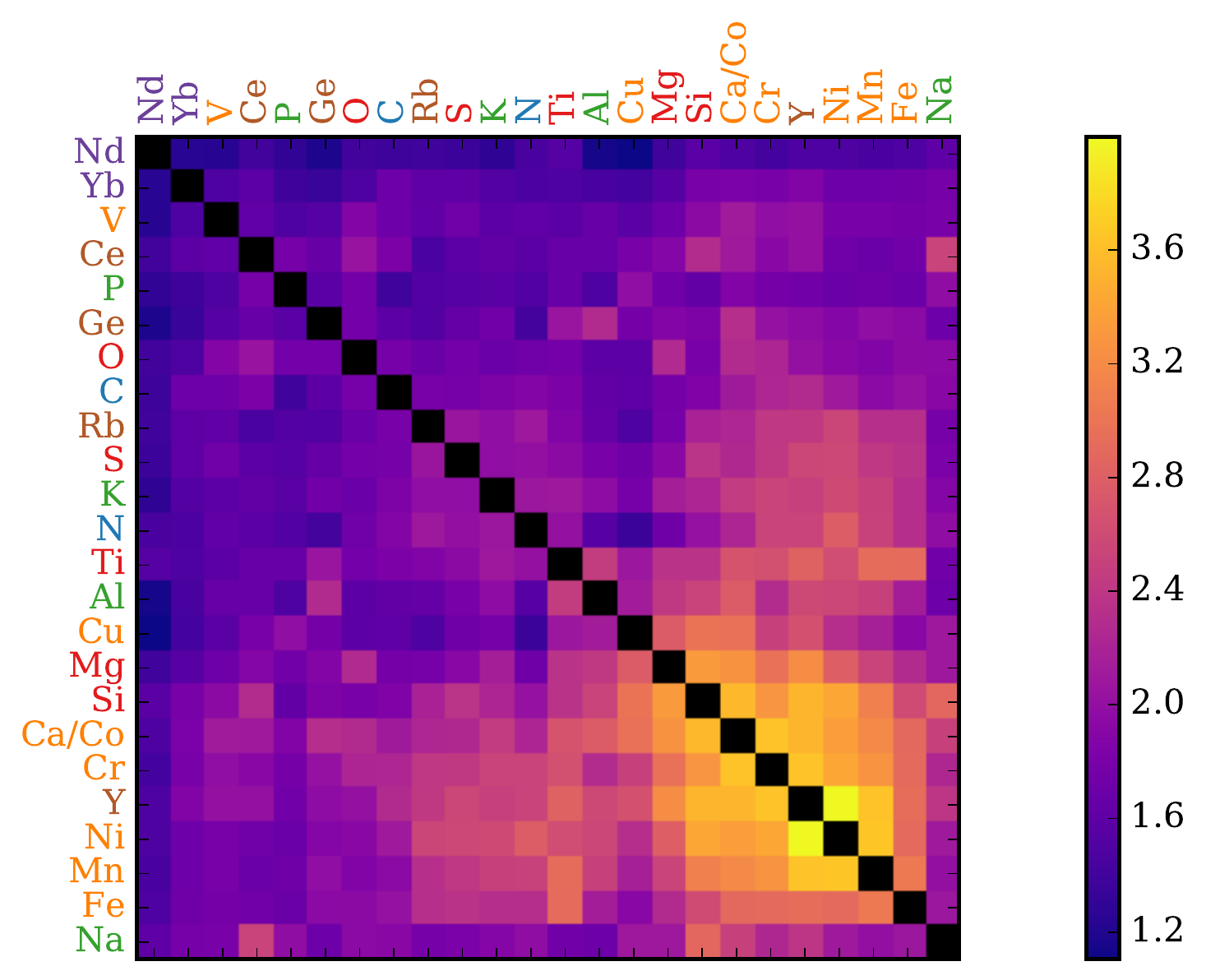}
    \caption{Information gains for pairs of elemental windows, as in the right panel of  Figure~\ref{fig:all_element_information} (with elements grouped by similar information content), now using 5 \AA\ windows in place of our standard 3 \AA\ windows. This Figure demonstrates the impact of the precise window definitions on the information gain: the magnitude of the gains has increased and the ordering of the windows has changed, though the iron-peak and alpha elements remain most predictive and grouped as before.}
    \label{fig:all_element_information_wide}
\end{figure}

As cautioned above, all of the conclusions reached thus far are conditional on the precise definitions of the elemental windows set out in Table~\ref{tab:window_centres}. To gain an impression of how generic these conclusions are, we repeat the above analysis using broader, 5 \AA\ windows, presenting a version of Figure~\ref{fig:all_element_information} for these windows in Figure~\ref{fig:all_element_information_wide}. There are numerous notes to make on this Figure. First, the scale extends to larger information gains: these windows are broader, contain more features and are therefore more predictive. The choice of first element that minimizes the total distance between rows in the plot is now neodymium, not nickel, but the structure is still similar: the most informative elements are from the iron-peak and alpha group, and these elements' similar predictivities mean they cluster in the plot. There is, again, something of a drop in information gains at nitrogen; however, the iron-peak and alpha elements now predict the other families better than before. Somewhat surprisingly, for these windows Y-Ni is the most informative pair. This is, however, due to an iron line (at around 15626 \AA) that appears in the yttrium window when it is extended to 5 \AA. Finally, note that increasing the bandwidth to 5 \AA\ causes our cobalt and calcium windows to merge.
These two last points serve to highlight again the fact that our conclusions derive from and apply to the full spectrum within each window, not necessarily solely to the element whose line defines the window centre. Careful consideration should be made of how to define and label windows in future work.

\section{Discussion \& Conclusions}
\label{sec:discussion}

In this work, we have demonstrated how to pool information from ensembles of stellar spectra in order to denoise and inpaint individual observations, introducing a method we call SSSpaNG. This has been done with the goal of optimizing the quality and quantity of measurements that can be made from stellar spectra, including chemical abundances and ages, from upcoming million-star spectroscopic surveys. We do so by modeling the distribution of 29502 APOGEE red clump stars' spectra as a high-dimensional Gaussian Process whose covariance matrix describes the variations in spectra within the population. Inferring the elements of this covariance matrix directly, we have shown that this completely data-driven model is capable of capturing the correlations between spectral pixels. These correlations can be harnessed to yield improved estimates of individual spectra, along with precise {\it and} accurate predictions for unobserved spectral pixels. By marginalizing over the covariance, we effectively place a non-Gaussian, highly sparsifying prior on these inferred spectra, strongly preferring spectra close to the population mean, but penalizing large deviations only logarithmically. We produce complete spectra with decreased uncertainties for each member of the population (reducing flux errors by a factor of 2-3 for stars with SNR $\approx$ 20). This denoising will enable improved abundance inference precision, for all elements, for every star. In particular, this provides significant opportunity for far higher fidelity abundance determinations for low SNR spectra. Our method therefore significantly enhances the precision of abundance estimation from data in hand. Equivalently, it suggests that precision abundance estimates can be achieved with less telescope time per spectrum.

We have demonstrated our method's potential using the recently discovered 15789 \AA\ cerium line, a high-value APOGEE target due to its s-process provenance~\citep{Cunha2017}. Our model makes accurate and precise predictions for this line in low-SNR stars in which the line is completely masked, permitting confident estimates of cerium abundances where they would previously have not been possible.

Modeling the red clump stars' spectra as a Gaussian Process also allows us to quantify the information gained by observing portions of a star's spectrum, and thereby define the most mutually informative regions of spectra. We have done so for windows centred on 25 elemental absorption lines in the APOGEE wavelength range, demonstrating that the iron-peak and alpha-process elements are particularly mutually informative. Harnessing this information, we are able to predict the spectrum in all but one of our example windows with uncertainty less than the APOGEE noise given high-precision observations of the single most-informative window. While we are unable to perfectly predict the flux in any single elemental window by observing a combination of other windows, we find that the majority of information about a target window is typically contained in the 10-or-so most informative windows. This is a clear demonstration of the power of using the data themselves to drive our understanding of the diversity of (and relationships between) different nucleosynthetic channels. Indeed, the correlation structure and information content that we can measure directly should place strong constraints on the physical processes that control chemical evolution. These relationships could inspire new, data-driven approaches to chemical evolution modeling \citep[also see][]{Casey2019}, replacing current theoretical approaches that fail to reproduce observed elemental yields in detail \citep[e.g.][]{Jan2017, Blancato2019}. Our information gain results also have important repercussions for the design of future observations, motivating the targeting of carefully selected, restricted spectral windows that yield strong predictions on a range of unobserved elements.

It is important at this point to address the current limitations of this method. The computational cost of the method is dominated by the matrix inversions required, which scale as the number of spectral pixels cubed. For each iteration of the Gibbs sampler we must perform one inversion per star and one inversion per class: too many for us to process the complete APOGEE dataset given available resources. In this work, we have restricted ourselves to narrow windows around our target elements; however, our results (most notably Figures~\ref{fig:single_element_information} and~\ref{fig:all_element_information_wide}) clearly show that there is significant value in including more of the spectra if possible. There are two obvious ways to achieve this: by throwing greater computational resources at the problem, or by exploiting the decaying eigenspectrum of the covariance matrices we find to infer a low-rank approximation to the covariance.

In the first approach, we can exploit the manifest parallelism in our algorithm. With access to the same number of CPUs as stars in the sample one could reduce the number of inversions per CPU per Gibbs sample to two at most.\footnote{We must invert each class's covariance matrix in order to sample the class memberships and true stellar spectra. We must also invert the sum of each star's inverse class covariance matrix and inverse noise covariance matrix in order to update its true spectrum. While we can parallelize the loops over classes and stars, the loops must be carried out sequentially, and thus some CPUs will always perform two inversions. If multiple CPUs were available for each star these inversions could also be parallelized, further reducing walltime.} Walltimes for our current 343-pixel runs are roughly 7 hours on 48 Intel Xeon CPUs; with 29502 CPUs the full dataset could therefore be processed in 7.5 days, though RAM-usage considerations might also affect this calculation. While clearly computationally heavy, this is feasible on existing large computing facilities.

In the second approach, the simplest way to reduce the rank of the inferred signal covariance matrix is to project the data onto the largest $m < \nb$ principal components of the {\em sample} covariance matrix prior to inference. Unfortunately, as the sample covariance contains both noise and signal its principal components are suboptimal for this task, severely degrading the inference. A natural solution would be to amend our model to explore only covariance matrices with a restricted structure (e.g., diagonal plus low-rank, along the lines of \citet{Zhang_etal:2013}). We leave such extensions to future work.\footnote{The structure of the covariance matrix also implies that certain kernels could potentially serve as useful covariance functions. Exploration of the utility of, for example, rational quadratic, Gibbs or mixtures of covariance functions~\citep{Rasmussen_Williams} is also left to future work.}

In the meantime, we are restricted to carrying out the analysis in windows as in this work. As the results depend entirely on the windows selected, the set of windows should be carefully optimized for the task at hand. In this proof-of-concept paper, we simply selected the strongest well-defined lines for a range of interesting elements, using a fixed bandwidth for all windows. For targeted applications, our information gain metric provides a well-motivated tool with which to optimize both the positions and widths of the elemental windows used. We have demonstrated in this work that restricting to a subset of windows still permits significant denoising and inpainting. This performance can be adapted to particular goals through careful definition of the windows; however, cutting the spectra clearly penalizes our ability to make serendipitous discoveries of new lines. We have shown here the method's ability to discover weak lines in noisy and masked spectra, but this is only possible because some stars have observed the relevant wavelengths. The loss of discovery space is a cost that must be weighed against improved performance in future applications of this work.

The final current limitation of this method is the poor sampling performance we observe when inferring the properties of multiple populations in our very high-dimensional APOGEE data. For the moment, we have chosen to model the red clump with a single class, asserting that the stars' binned spectra are distributed as a multivariate normal. As such, our handling of contaminants (or outliers) is suboptimal. Contaminants will manifest as non-Gaussianity or multi-modality in the bulk population, and will therefore increase the variance of the inferred true spectra and covariance matrix if incorrectly modeled as a single Gaussian population. We do not expect contaminants to impact our results strongly, as they are estimated to make up only 5-10\% of our red clump sample \citep{Bovy2014}, but the same can not be said for more diverse datasets. We know that different stellar populations have different spectral correlation structures: globular clusters, for example, have known abundance anti-correlations that are not seen in the disk and field halo stars~\citep[e.g.,][]{Kraft1997,Gratton2015, Pan2017,Carr2019}. Demonstrating that our sampler can efficiently and accurately fit multiple classes will allow us to not only model datasets containing different, potentially non-Gaussian populations completely, but also discover new populations. This is particularly interesting as it ties into, for example, a method of understanding chemodynamical classes in the Galactic halo, which is expected to consist of discrete chemical sub-systems with different elemental correlations. As with the other limitations, investigating modifications to the sampler (simulated annealing, for example) to address this issue, is left to future work.


\section*{Acknowledgements}

The Flatiron Institute is supported by the Simons Foundation.
SMF is supported by the Royal Society. MKN is supported in part by the Sloan Foundation. We would like to thank Brice Menard (JHU) who was instrumental in bringing our team together to perform this work. 


\section*{Data Availability}

The data employed in this article are available for download from the Sloan Digital Sky Survey's Value Added Catalogs, at \url{https://www.sdss.org/dr14/data_access/value-added-catalogs/?vac_id=apogee-red-clump-rc-catalog}.


\bibliographystyle{mnras}
\bibliography{mknbib}

\begin{thebibliography}{}
\makeatletter
\relax
\def\mn@urlcharsother{\let\do\@makeother \do\$\do\&\do\#\do\^\do\_\do\%\do\~}
\def\mn@doi{\begingroup\mn@urlcharsother \@ifnextchar [ {\mn@doi@}
  {\mn@doi@[]}}
\def\mn@doi@[#1]#2{\def\@tempa{#1}\ifx\@tempa\@empty \href
  {http://dx.doi.org/#2} {doi:#2}\else \href {http://dx.doi.org/#2} {#1}\fi
  \endgroup}
\def\mn@eprint#1#2{\mn@eprint@#1:#2::\@nil}
\def\mn@eprint@arXiv#1{\href {http://arxiv.org/abs/#1} {{\tt arXiv:#1}}}
\def\mn@eprint@dblp#1{\href {http://dblp.uni-trier.de/rec/bibtex/#1.xml}
  {dblp:#1}}
\def\mn@eprint@#1:#2:#3:#4\@nil{\def\@tempa {#1}\def\@tempb {#2}\def\@tempc
  {#3}\ifx \@tempc \@empty \let \@tempc \@tempb \let \@tempb \@tempa \fi \ifx
  \@tempb \@empty \def\@tempb {arXiv}\fi \@ifundefined
  {mn@eprint@\@tempb}{\@tempb:\@tempc}{\expandafter \expandafter \csname
  mn@eprint@\@tempb\endcsname \expandafter{\@tempc}}}

\bibitem[\protect\citeauthoryear{{Armillotta}, {Krumholz}  \&
  {Fujimoto}}{{Armillotta} et~al.}{2018}]{Arm2018}
{Armillotta} L.,  {Krumholz} M.~R.,   {Fujimoto} Y.,  2018, \mn@doi [\mnras]
  {10.1093/mnras/sty2625}, \href
  {https://ui.adsabs.harvard.edu/abs/2018MNRAS.481.5000A} {481, 5000}

\bibitem[\protect\citeauthoryear{{Blancato}, {Ness}, {Johnston}, {Rybizki}  \&
  {Bedell}}{{Blancato} et~al.}{2019}]{Blancato2019}
{Blancato} K.,  {Ness} M.,  {Johnston} K.~V.,  {Rybizki} J.,   {Bedell} M.,
  2019, arXiv e-prints, \href
  {https://ui.adsabs.harvard.edu/abs/2019arXiv190605297B} {p. arXiv:1906.05297}

\bibitem[\protect\citeauthoryear{{Bland-Hawthorn} \&
  {Gerhard}}{{Bland-Hawthorn} \& {Gerhard}}{2016}]{BH2016}
{Bland-Hawthorn} J.,  {Gerhard} O.,  2016, \mn@doi [\araa]
  {10.1146/annurev-astro-081915-023441}, \href
  {https://ui.adsabs.harvard.edu/abs/2016ARA&A..54..529B} {54, 529}

\bibitem[\protect\citeauthoryear{{Bland-Hawthorn}, {Krumholz}  \&
  {Freeman}}{{Bland-Hawthorn} et~al.}{2010}]{BH2010}
{Bland-Hawthorn} J.,  {Krumholz} M.~R.,   {Freeman} K.,  2010, \mn@doi [\apj]
  {10.1088/0004-637X/713/1/166}, \href
  {http://adsabs.harvard.edu/abs/2010ApJ...713..166B} {713, 166}

\bibitem[\protect\citeauthoryear{{Bland-Hawthorn} et~al.,}{{Bland-Hawthorn}
  et~al.}{2019}]{BH2019}
{Bland-Hawthorn} J.,  et~al., 2019, \mn@doi [\mnras] {10.1093/mnras/stz217},
  \href {http://adsabs.harvard.edu/abs/2019MNRAS.tmp..222B} {}

\bibitem[\protect\citeauthoryear{{Bond} \& {Efstathiou}}{{Bond} \&
  {Efstathiou}}{1987}]{Bond_etal:1987}
{Bond} J.~R.,  {Efstathiou} G.,  1987, \mn@doi [\mnras]
  {10.1093/mnras/226.3.655}, \href
  {https://ui.adsabs.harvard.edu/abs/1987MNRAS.226..655B} {226, 655}

\bibitem[\protect\citeauthoryear{{Bonifacio} et~al.,}{{Bonifacio}
  et~al.}{2016}]{Bonifacio2016}
{Bonifacio} P.,  et~al., 2016, in {Reyl{\'e}} C.,  {Richard} J.,
  {Cambr{\'e}sy} L.,  {Deleuil} M.,  {P{\'e}contal} E.,  {Tresse} L.,
  {Vauglin} I.,  eds, SF2A-2016: Proceedings of the Annual meeting of the
  French Society of Astronomy and Astrophysics. pp 267--270

\bibitem[\protect\citeauthoryear{{Bovy}, {Nidever}, {Rix}, {Girardi},
  {Zasowski}, {Chojnowski}  \& {Holtzman}}{{Bovy} et~al.}{2014}]{Bovy2014}
{Bovy} J.,  {Nidever} D.~L.,  {Rix} H.-W.,  {Girardi} L.,  {Zasowski} G.,
  {Chojnowski} S.~D.,   {Holtzman} J. e.~a.,  2014, \mn@doi [\apj]
  {10.1088/0004-637X/790/2/127}, \href
  {http://adsabs.harvard.edu/abs/2014ApJ...790..127B} {790, 127}

\bibitem[\protect\citeauthoryear{{Bovy}, {Leung}, {Hunt}, {Mackereth},
  {Garcia-Hernandez}  \& {Roman-Lopes}}{{Bovy} et~al.}{2019}]{Bovy2019}
{Bovy} J.,  {Leung} H.~W.,  {Hunt} J. A.~S.,  {Mackereth} J.~T.,
  {Garcia-Hernandez} D.~A.,   {Roman-Lopes} A.,  2019, arXiv e-prints, \href
  {https://ui.adsabs.harvard.edu/abs/2019arXiv190511404B} {p. arXiv:1905.11404}

\bibitem[\protect\citeauthoryear{Busemeyer, Wang, Townsend  \&
  Eidels}{Busemeyer et~al.}{2015}]{Busemeyer_etal:2015}
Busemeyer J.~R.,  Wang Z.,  Townsend J.~T.,   Eidels A.,  2015, The Oxford
  Handbook of Computational and Mathematical Psychology.
Oxford University Press, \url
  {https://www.oxfordhandbooks.com/view/10.1093/oxfordhb/9780199957996.001.0001/oxfordhb-9780199957996}

\bibitem[\protect\citeauthoryear{{Carretta}}{{Carretta}}{2019}]{Carr2019}
{Carretta} E.,  2019, \mn@doi [\aap] {10.1051/0004-6361/201935110}, \href
  {https://ui.adsabs.harvard.edu/abs/2019A&A...624A..24C} {624, A24}

\bibitem[\protect\citeauthoryear{{Casey}, {Hogg}, {Ness}, {Rix}, {Ho}  \&
  {Gilmore}}{{Casey} et~al.}{2016}]{Casey2016}
{Casey} A.~R.,  {Hogg} D.~W.,  {Ness} M.,  {Rix} H.-W.,  {Ho} A.~Q.,
  {Gilmore} G.,  2016, preprint, \href
  {http://adsabs.harvard.edu/abs/2016arXiv160303040C} {} (\mn@eprint {arXiv}
  {1603.03040})

\bibitem[\protect\citeauthoryear{{Casey} et~al.,}{{Casey}
  et~al.}{2019}]{Casey2019}
{Casey} A.~R.,  et~al., 2019, arXiv e-prints, \href
  {https://ui.adsabs.harvard.edu/abs/2019arXiv191009811C} {p. arXiv:1910.09811}

\bibitem[\protect\citeauthoryear{{Cirasuolo}, {Afonso}, {Carollo}, {Flores}  \&
  {Maiolino}}{{Cirasuolo} et~al.}{2014}]{C2014}
{Cirasuolo} M.,  {Afonso} J.,  {Carollo} M.,  {Flores} H.,   {Maiolino} R.
  e.~a.,  2014, in Ground-based and Airborne Instrumentation for Astronomy V.
  p. 91470N, \mn@doi{10.1117/12.2056012}

\bibitem[\protect\citeauthoryear{{Clarke} et~al.,}{{Clarke}
  et~al.}{2019}]{Clarke2019}
{Clarke} A.~J.,  et~al., 2019, \mn@doi [\mnras] {10.1093/mnras/stz104}, \href
  {http://adsabs.harvard.edu/abs/2019MNRAS.484.3476C} {484, 3476}

\bibitem[\protect\citeauthoryear{Cover \& Thomas}{Cover \&
  Thomas}{2006}]{Cover_Thomas:2006}
Cover T.~M.,  Thomas J.~A.,  2006, Elements of Information Theory (Wiley Series
  in Telecommunications and Signal Processing).
Wiley-Interscience, New York, NY, USA

\bibitem[\protect\citeauthoryear{{Cunha} et~al.,}{{Cunha}
  et~al.}{2017}]{Cunha2017}
{Cunha} K.,  et~al., 2017, \mn@doi [\apj] {10.3847/1538-4357/aa7beb}, \href
  {http://adsabs.harvard.edu/abs/2017ApJ...844..145C} {844, 145}

\bibitem[\protect\citeauthoryear{{Czekala}, {Mandel}, {Andrews}, {Dittmann},
  {Ghosh}, {Montet}  \& {Newton}}{{Czekala} et~al.}{2017}]{Czekala_etal:2017}
{Czekala} I.,  {Mandel} K.~S.,  {Andrews} S.~M.,  {Dittmann} J.~A.,  {Ghosh}
  S.~K.,  {Montet} B.~T.,   {Newton} E.~R.,  2017, \mn@doi [\apj]
  {10.3847/1538-4357/aa6aab}, \href
  {https://ui.adsabs.harvard.edu/abs/2017ApJ...840...49C} {840, 49}

\bibitem[\protect\citeauthoryear{{Das}, {Hawkins}  \& {Jofre}}{{Das}
  et~al.}{2019}]{Payel2019}
{Das} P.,  {Hawkins} K.,   {Jofre} P.,  2019, arXiv e-prints, \href
  {https://ui.adsabs.harvard.edu/abs/2019arXiv190309320D} {p. arXiv:1903.09320}

\bibitem[\protect\citeauthoryear{{De Silva} et~al.,}{{De Silva}
  et~al.}{2015}]{deSilva2015}
{De Silva} G.~M.,  et~al., 2015, \mn@doi [\mnras] {10.1093/mnras/stv327}, \href
  {http://adsabs.harvard.edu/abs/2015MNRAS.449.2604D} {449, 2604}

\bibitem[\protect\citeauthoryear{{Foreman-Mackey}, {Hogg}  \&
  {Morton}}{{Foreman-Mackey} et~al.}{2014}]{DFM_etal:2014}
{Foreman-Mackey} D.,  {Hogg} D.~W.,   {Morton} T.~D.,  2014, \mn@doi [\apj]
  {10.1088/0004-637X/795/1/64}, \href
  {https://ui.adsabs.harvard.edu/abs/2014ApJ...795...64F} {795, 64}

\bibitem[\protect\citeauthoryear{{Frankel}, {Rix}, {Ting}, {Ness}  \&
  {Hogg}}{{Frankel} et~al.}{2018}]{Frankel2018}
{Frankel} N.,  {Rix} H.-W.,  {Ting} Y.-S.,  {Ness} M.~K.,   {Hogg} D.~W.,
  2018, preprint, \href {http://adsabs.harvard.edu/abs/2018arXiv180509198F} {}
  (\mn@eprint {arXiv} {1805.09198})

\bibitem[\protect\citeauthoryear{{Gaia Collaboration} et~al.,}{{Gaia
  Collaboration} et~al.}{2016}]{Gaia2016}
{Gaia Collaboration} et~al., 2016, \mn@doi [\aap]
  {10.1051/0004-6361/201629512}, \href
  {http://adsabs.harvard.edu/abs/2016A%26A...595A...2G} {595, A2}

\bibitem[\protect\citeauthoryear{{Garc{\'{\i}}a P{\'e}rez}, {Allende Prieto},
  {Holtzman}, {Shetrone}  \& {M{\'e}sz{\'a}ros}}{{Garc{\'{\i}}a P{\'e}rez}
  et~al.}{2015}]{GP2015}
{Garc{\'{\i}}a P{\'e}rez} A.~E.,  {Allende Prieto} C.,  {Holtzman} J.~A.,
  {Shetrone} M.,   {M{\'e}sz{\'a}ros} S. e.~a.,  2015, preprint, \href
  {http://adsabs.harvard.edu/abs/2015arXiv151007635G} {} (\mn@eprint {arXiv}
  {1510.07635})

\bibitem[\protect\citeauthoryear{Gelman, Carlin, Stern  \& Rubin}{Gelman
  et~al.}{2013}]{Gelman_etal:2013}
Gelman A.,  Carlin J.~B.,  Stern H.~S.,   Rubin D.~B.,  2013, Bayesian Data
  Analysis, 3rd edn.
Chapman and Hall/CRC

\bibitem[\protect\citeauthoryear{{Geman} \& {Geman}}{{Geman} \&
  {Geman}}{1984}]{Geman_and_Geman:1984}
{Geman} S.,  {Geman} D.,  1984, \mn@doi [IEEE Transactions on Pattern Analysis
  and Machine Intelligence] {10.1109/TPAMI.1984.4767596}, PAMI-6, 721

\bibitem[\protect\citeauthoryear{{Gibson}, {Aigrain}, {Roberts}, {Evans},
  {Osborne}  \& {Pont}}{{Gibson} et~al.}{2012}]{Gibson_etal:2012}
{Gibson} N.~P.,  {Aigrain} S.,  {Roberts} S.,  {Evans} T.~M.,  {Osborne} M.,
  {Pont} F.,  2012, \mn@doi [\mnras] {10.1111/j.1365-2966.2011.19915.x}, \href
  {https://ui.adsabs.harvard.edu/abs/2012MNRAS.419.2683G} {419, 2683}

\bibitem[\protect\citeauthoryear{{Gilmore} et~al.,}{{Gilmore}
  et~al.}{2012}]{Gilmore2012}
{Gilmore} G.,  et~al., 2012, The Messenger, \href
  {http://adsabs.harvard.edu/abs/2012Msngr.147...25G} {147, 25}

\bibitem[\protect\citeauthoryear{{Gratton} et~al.,}{{Gratton}
  et~al.}{2015}]{Gratton2015}
{Gratton} R.~G.,  et~al., 2015, \mn@doi [\aap] {10.1051/0004-6361/201424393},
  \href {https://ui.adsabs.harvard.edu/abs/2015A&A...573A..92G} {573, A92}

\bibitem[\protect\citeauthoryear{{Hasselquist} et~al.,}{{Hasselquist}
  et~al.}{2016}]{Hasselquist_etal:2016}
{Hasselquist} S.,  et~al., 2016, \mn@doi [\apj] {10.3847/1538-4357/833/1/81},
  \href {https://ui.adsabs.harvard.edu/abs/2016ApJ...833...81H} {833, 81}

\bibitem[\protect\citeauthoryear{Hastings}{Hastings}{1970}]{Hastings:1970}
Hastings W.~K.,  1970, \mn@doi [Biometrika] {10.1093/biomet/57.1.97}, 57, 97

\bibitem[\protect\citeauthoryear{{Hawkins}, {Jofr{\'e}}, {Masseron}  \&
  {Gilmore}}{{Hawkins} et~al.}{2015}]{Keith2015}
{Hawkins} K.,  {Jofr{\'e}} P.,  {Masseron} T.,   {Gilmore} G.,  2015, \mn@doi
  [\mnras] {10.1093/mnras/stv1586}, \href
  {https://ui.adsabs.harvard.edu/abs/2015MNRAS.453..758H} {453, 758}

\bibitem[\protect\citeauthoryear{{Hayden} et~al.,}{{Hayden}
  et~al.}{2015}]{Hayden2015}
{Hayden} M.~R.,  et~al., 2015, \mn@doi [\apj] {10.1088/0004-637X/808/2/132},
  \href {http://adsabs.harvard.edu/abs/2015ApJ...808..132H} {808, 132}

\bibitem[\protect\citeauthoryear{{Helmi}, {Babusiaux}, {Koppelman}, {Massari},
  {Veljanoski}  \& {Brown}}{{Helmi} et~al.}{2018}]{Helmi2018}
{Helmi} A.,  {Babusiaux} C.,  {Koppelman} H.~H.,  {Massari} D.,  {Veljanoski}
  J.,   {Brown} A. G.~A.,  2018, \mn@doi [\nat] {10.1038/s41586-018-0625-x},
  \href {https://ui.adsabs.harvard.edu/abs/2018Natur.563...85H} {563, 85}

\bibitem[\protect\citeauthoryear{{Ho}, {Rix}, {Ness}, {Hogg}, {Liu}  \&
  {Ting}}{{Ho} et~al.}{2017a}]{Ho2017b}
{Ho} A.~Y.~Q.,  {Rix} H.-W.,  {Ness} M.~K.,  {Hogg} D.~W.,  {Liu} C.,   {Ting}
  Y.-S.,  2017a, \mn@doi [\apj] {10.3847/1538-4357/aa6db3}, \href
  {http://adsabs.harvard.edu/abs/2017ApJ...841...40H} {841, 40}

\bibitem[\protect\citeauthoryear{{Ho}, {Rix}, {Ness}, {Hogg}, {Liu}  \&
  {Ting}}{{Ho} et~al.}{2017b}]{Ho2017}
{Ho} A. Y.~Q.,  {Rix} H.-W.,  {Ness} M.~K.,  {Hogg} D.~W.,  {Liu} C.,   {Ting}
  Y.-S.,  2017b, \mn@doi [\apj] {10.3847/1538-4357/aa6db3}, \href
  {https://ui.adsabs.harvard.edu/abs/2017ApJ...841...40H} {841, 40}

\bibitem[\protect\citeauthoryear{{Hogg} et~al.,}{{Hogg}
  et~al.}{2016}]{Hogg2016}
{Hogg} D.~W.,  et~al., 2016, \mn@doi [\apj] {10.3847/1538-4357/833/2/262},
  \href {https://ui.adsabs.harvard.edu/abs/2016ApJ...833..262H} {833, 262}

\bibitem[\protect\citeauthoryear{{Holtzman} et~al.,}{{Holtzman}
  et~al.}{2015}]{Holtzman2015}
{Holtzman} J.~A.,  et~al., 2015, \mn@doi [\aj] {10.1088/0004-6256/150/5/148},
  \href {http://adsabs.harvard.edu/abs/2015AJ....150..148H} {150, 148}

\bibitem[\protect\citeauthoryear{{Kollmeier} et~al.,}{{Kollmeier}
  et~al.}{2017}]{Kollmeier2017}
{Kollmeier} J.~A.,  et~al., 2017, preprint, \href
  {http://adsabs.harvard.edu/abs/2017arXiv171103234K} {} (\mn@eprint {arXiv}
  {1711.03234})

\bibitem[\protect\citeauthoryear{{Kordopatis} et~al.,}{{Kordopatis}
  et~al.}{2015}]{Kord2015}
{Kordopatis} G.,  et~al., 2015, \mn@doi [\aap] {10.1051/0004-6361/201526258},
  \href {https://ui.adsabs.harvard.edu/abs/2015A&A...582A.122K} {582, A122}

\bibitem[\protect\citeauthoryear{{Kraft}, {Sneden}, {Smith}, {Shetrone},
  {Langer}  \& {Pilachowski}}{{Kraft} et~al.}{1997}]{Kraft1997}
{Kraft} R.~P.,  {Sneden} C.,  {Smith} G.~H.,  {Shetrone} M.~D.,  {Langer}
  G.~E.,   {Pilachowski} C.~A.,  1997, \mn@doi [\aj] {10.1086/118251}, \href
  {https://ui.adsabs.harvard.edu/abs/1997AJ....113..279K} {113, 279}

\bibitem[\protect\citeauthoryear{{Leung} \& {Bovy}}{{Leung} \&
  {Bovy}}{2019}]{Leung2019}
{Leung} H.~W.,  {Bovy} J.,  2019, \mn@doi [\mnras] {10.1093/mnras/sty3217},
  \href {https://ui.adsabs.harvard.edu/abs/2019MNRAS.483.3255L} {483, 3255}

\bibitem[\protect\citeauthoryear{{Mackereth} et~al.,}{{Mackereth}
  et~al.}{2019}]{Ted2019a}
{Mackereth} J.~T.,  et~al., 2019, arXiv e-prints, \href
  {http://adsabs.harvard.edu/abs/2019arXiv190104502M} {}

\bibitem[\protect\citeauthoryear{{Majewski} et~al.,}{{Majewski}
  et~al.}{2017}]{Majewski2017}
{Majewski} S.~R.,  et~al., 2017, \mn@doi [\aj] {10.3847/1538-3881/aa784d},
  \href {http://adsabs.harvard.edu/abs/2017AJ....154...94M} {154, 94}

\bibitem[\protect\citeauthoryear{{Minchev}, {Chiappini}  \& {Martig}}{{Minchev}
  et~al.}{2013}]{Minchev2013}
{Minchev} I.,  {Chiappini} C.,   {Martig} M.,  2013, \mn@doi [\aap]
  {10.1051/0004-6361/201220189}, \href
  {https://ui.adsabs.harvard.edu/abs/2013A&A...558A...9M} {558, A9}

\bibitem[\protect\citeauthoryear{{Minchev}, {Chiappini}  \& {Martig}}{{Minchev}
  et~al.}{2014a}]{Minchev2014b}
{Minchev} I.,  {Chiappini} C.,   {Martig} M.,  2014a, \mn@doi [\aap]
  {10.1051/0004-6361/201423487}, \href
  {https://ui.adsabs.harvard.edu/abs/2014A&A...572A..92M} {572, A92}

\bibitem[\protect\citeauthoryear{{Minchev} et~al.,}{{Minchev}
  et~al.}{2014b}]{Minchev2014a}
{Minchev} I.,  et~al., 2014b, \mn@doi [\apj] {10.1088/2041-8205/781/1/L20},
  \href {https://ui.adsabs.harvard.edu/abs/2014ApJ...781L..20M} {781, L20}

\bibitem[\protect\citeauthoryear{{Mitschang}, {De Silva}, {Sharma}  \&
  {Zucker}}{{Mitschang} et~al.}{2013}]{Mits2013}
{Mitschang} A.~W.,  {De Silva} G.,  {Sharma} S.,   {Zucker} D.~B.,  2013,
  \mn@doi [\mnras] {10.1093/mnras/sts194}, \href
  {https://ui.adsabs.harvard.edu/abs/2013MNRAS.428.2321M} {428, 2321}

\bibitem[\protect\citeauthoryear{{Mitschang}, {De Silva}, {Zucker}, {Anguiano},
  {Bensby}  \& {Feltzing}}{{Mitschang} et~al.}{2014}]{M2014}
{Mitschang} A.~W.,  {De Silva} G.,  {Zucker} D.~B.,  {Anguiano} B.,  {Bensby}
  T.,   {Feltzing} S.,  2014, \mn@doi [\mnras] {10.1093/mnras/stt2320}, \href
  {http://adsabs.harvard.edu/abs/2014MNRAS.438.2753M} {438, 2753}

\bibitem[\protect\citeauthoryear{{Ness}}{{Ness}}{2018}]{Ness2018}
{Ness} M.,  2018, \mn@doi [\pasa] {10.1017/pasa.2017.53}, \href
  {http://adsabs.harvard.edu/abs/2018PASA...35....3N} {35, e003}

\bibitem[\protect\citeauthoryear{{Ness}, {Hogg}, {Rix}, {Ho}  \&
  {Zasowski}}{{Ness} et~al.}{2015}]{Ness2015}
{Ness} M.,  {Hogg} D.~W.,  {Rix} H.-W.,  {Ho} A.~Y.~Q.,   {Zasowski} G.,  2015,
  \mn@doi [\apj] {10.1088/0004-637X/808/1/16}, \href
  {http://adsabs.harvard.edu/abs/2015ApJ...808...16N} {808, 16}

\bibitem[\protect\citeauthoryear{{Newberg} et~al.,}{{Newberg}
  et~al.}{2012}]{Newberg2012}
{Newberg} H.~J.,  et~al., 2012, in {Aoki} W.,  {Ishigaki} M.,  {Suda} T.,
  {Tsujimoto} T.,   {Arimoto} N.,  eds,  Astronomical Society of the Pacific
  Conference Series Vol. 458, Galactic Archaeology: Near-Field Cosmology and
  the Formation of the Milky Way. p.~405

\bibitem[\protect\citeauthoryear{{Nidever} et~al.,}{{Nidever}
  et~al.}{2014}]{Nidever2014}
{Nidever} D.~L.,  et~al., 2014, \mn@doi [\apj] {10.1088/0004-637X/796/1/38},
  \href {http://adsabs.harvard.edu/abs/2014ApJ...796...38N} {796, 38}

\bibitem[\protect\citeauthoryear{{Nidever} et~al.,}{{Nidever}
  et~al.}{2015}]{Nidever2015}
{Nidever} D.~L.,  et~al., 2015, \mn@doi [\aj] {10.1088/0004-6256/150/6/173},
  \href {https://ui.adsabs.harvard.edu/abs/2015AJ....150..173N} {150, 173}

\bibitem[\protect\citeauthoryear{{Pancino} et~al.,}{{Pancino}
  et~al.}{2017}]{Pan2017}
{Pancino} E.,  et~al., 2017, \mn@doi [\aap] {10.1051/0004-6361/201730474},
  \href {https://ui.adsabs.harvard.edu/abs/2017A&A...601A.112P} {601, A112}

\bibitem[\protect\citeauthoryear{{Price-Jones} \& {Bovy}}{{Price-Jones} \&
  {Bovy}}{2019}]{PJ2019}
{Price-Jones} N.,  {Bovy} J.,  2019, arXiv e-prints, \href
  {http://adsabs.harvard.edu/abs/2019arXiv190208201P} {}

\bibitem[\protect\citeauthoryear{Rasmussen \& Williams}{Rasmussen \&
  Williams}{2006}]{Rasmussen_Williams}
Rasmussen C.,  Williams C.,  2006, Gaussian Processes for Machine Learning.
Adaptive Computation and Machine Learning, MIT Press, Cambridge, MA, USA

\bibitem[\protect\citeauthoryear{{Rix} \& {Bovy}}{{Rix} \&
  {Bovy}}{2013}]{Rix2013}
{Rix} H.-W.,  {Bovy} J.,  2013, \mn@doi [\aapr] {10.1007/s00159-013-0061-8},
  \href {https://ui.adsabs.harvard.edu/abs/2013A&ARv..21...61R} {21, 61}

\bibitem[\protect\citeauthoryear{{Rybizki}, {Just}  \& {Rix}}{{Rybizki}
  et~al.}{2017}]{Jan2017}
{Rybizki} J.,  {Just} A.,   {Rix} H.-W.,  2017, \mn@doi [\aap]
  {10.1051/0004-6361/201730522}, \href
  {https://ui.adsabs.harvard.edu/abs/2017A&A...605A..59R} {605, A59}

\bibitem[\protect\citeauthoryear{{Sanderson} et~al.,}{{Sanderson}
  et~al.}{2018}]{Robyn2018}
{Sanderson} R.~E.,  et~al., 2018, arXiv e-prints, \href
  {https://ui.adsabs.harvard.edu/abs/2018arXiv180610564S} {p. arXiv:1806.10564}

\bibitem[\protect\citeauthoryear{{Shafieloo}, {Kim}  \& {Linder}}{{Shafieloo}
  et~al.}{2012}]{Shafieloo_etal:2012}
{Shafieloo} A.,  {Kim} A.~G.,   {Linder} E.~V.,  2012, \mn@doi [\prd]
  {10.1103/PhysRevD.85.123530}, \href
  {https://ui.adsabs.harvard.edu/abs/2012PhRvD..85l3530S} {85, 123530}

\bibitem[\protect\citeauthoryear{{Shetrone} et~al.,}{{Shetrone}
  et~al.}{2015}]{Shetrone2015}
{Shetrone} M.,  et~al., 2015, \mn@doi [\apjs] {10.1088/0067-0049/221/2/24},
  \href {http://adsabs.harvard.edu/abs/2015ApJS..221...24S} {221, 24}

\bibitem[\protect\citeauthoryear{{Steinmetz} et~al.,}{{Steinmetz}
  et~al.}{2006}]{Steinmetz2006}
{Steinmetz} M.,  et~al., 2006, \mn@doi [\aj] {10.1086/506564}, \href
  {http://adsabs.harvard.edu/abs/2006AJ....132.1645S} {132, 1645}

\bibitem[\protect\citeauthoryear{{Sutter} et~al.,}{{Sutter}
  et~al.}{2014}]{Sutter_etal:2014}
{Sutter} P.~M.,  et~al., 2014, \mn@doi [\mnras] {10.1093/mnras/stt2244}, \href
  {https://ui.adsabs.harvard.edu/abs/2014MNRAS.438..768S} {438, 768}

\bibitem[\protect\citeauthoryear{{Tamura} et~al.,}{{Tamura}
  et~al.}{2016}]{PFS2016}
{Tamura} N.,  et~al., 2016, in Ground-based and Airborne Instrumentation for
  Astronomy VI. p. 99081M (\mn@eprint {arXiv} {1608.01075}),
  \mn@doi{10.1117/12.2232103}

\bibitem[\protect\citeauthoryear{{Ting}, {Freeman}, {Kobayashi}, {De Silva}  \&
  {Bland-Hawthorn}}{{Ting} et~al.}{2012}]{Ting2012}
{Ting} Y.-S.,  {Freeman} K.~C.,  {Kobayashi} C.,  {De Silva} G.~M.,
  {Bland-Hawthorn} J.,  2012, \mn@doi [\mnras]
  {10.1111/j.1365-2966.2011.20387.x}, \href
  {https://ui.adsabs.harvard.edu/abs/2012MNRAS.421.1231T} {421, 1231}

\bibitem[\protect\citeauthoryear{{Ting}, {Conroy}  \& {Goodman}}{{Ting}
  et~al.}{2015}]{Ting2015}
{Ting} Y.-S.,  {Conroy} C.,   {Goodman} A.,  2015, \mn@doi [\apj]
  {10.1088/0004-637X/807/1/104}, \href
  {http://adsabs.harvard.edu/abs/2015ApJ...807..104T} {807, 104}

\bibitem[\protect\citeauthoryear{{Ting}, {Rix}, {Conroy}, {Ho}  \&
  {Lin}}{{Ting} et~al.}{2017}]{Ting2017}
{Ting} Y.-S.,  {Rix} H.-W.,  {Conroy} C.,  {Ho} A.~Y.~Q.,   {Lin} J.,  2017,
  preprint, \href {http://adsabs.harvard.edu/abs/2017arXiv170801758T} {}
  (\mn@eprint {arXiv} {1708.01758})

\bibitem[\protect\citeauthoryear{{Ting}, {Conroy}, {Rix}  \& {Cargile}}{{Ting}
  et~al.}{2018}]{Ting2018}
{Ting} Y.-S.,  {Conroy} C.,  {Rix} H.-W.,   {Cargile} P.,  2018, arXiv
  e-prints, \href {https://ui.adsabs.harvard.edu/abs/2018arXiv180401530T} {p.
  arXiv:1804.01530}

\bibitem[\protect\citeauthoryear{{Weinberg} et~al.,}{{Weinberg}
  et~al.}{2019}]{Weinberg2019}
{Weinberg} D.~H.,  et~al., 2019, \mn@doi [\apj] {10.3847/1538-4357/ab07c7},
  \href {http://adsabs.harvard.edu/abs/2019ApJ...874..102W} {874, 102}

\bibitem[\protect\citeauthoryear{{Xiang} et~al.,}{{Xiang}
  et~al.}{2019}]{Xiang2019}
{Xiang} M.,  et~al., 2019, \mn@doi [\apjs] {10.3847/1538-4365/ab5364}, \href
  {https://ui.adsabs.harvard.edu/abs/2019ApJS..245...34X} {245, 34}

\bibitem[\protect\citeauthoryear{{Yanny} et~al.,}{{Yanny}
  et~al.}{2009}]{Yanny2009}
{Yanny} B.,  et~al., 2009, \mn@doi [\aj] {10.1088/0004-6256/137/5/4377}, \href
  {http://adsabs.harvard.edu/abs/2009AJ....137.4377Y} {137, 4377}

\bibitem[\protect\citeauthoryear{{Zasowski} et~al.,}{{Zasowski}
  et~al.}{2013}]{Zasowski2013}
{Zasowski} G.,  et~al., 2013, \mn@doi [\aj] {10.1088/0004-6256/146/4/81}, \href
  {https://ui.adsabs.harvard.edu/abs/2013AJ....146...81Z} {146, 81}

\bibitem[\protect\citeauthoryear{{Zasowski} et~al.,}{{Zasowski}
  et~al.}{2017}]{Zasowski2017}
{Zasowski} G.,  et~al., 2017, \mn@doi [\aj] {10.3847/1538-3881/aa8df9}, \href
  {https://ui.adsabs.harvard.edu/abs/2017AJ....154..198Z} {154, 198}

\bibitem[\protect\citeauthoryear{{Zhang}, {Sarkar}  \& {Mallick}}{{Zhang}
  et~al.}{2013}]{Zhang_etal:2013}
{Zhang} L.,  {Sarkar} A.,   {Mallick} B.~K.,  2013, arXiv e-prints, \href
  {https://ui.adsabs.harvard.edu/abs/2013arXiv1310.4195Z} {p. arXiv:1310.4195}

\bibitem[\protect\citeauthoryear{{de Jong} et~al.,}{{de Jong}
  et~al.}{2016}]{deJong2016}
{de Jong} R.~S.,  et~al., 2016, in Ground-based and Airborne Instrumentation
  for Astronomy VI. p. 99081O, \mn@doi{10.1117/12.2232832}

\makeatother
\end{thebibliography}


\bsp	
\label{lastpage}
\end{document}